\documentclass[journal]{IEEEtran}
\IEEEoverridecommandlockouts
\usepackage{lineno}
\usepackage{url}

\usepackage{cite}
\usepackage{amsmath,amssymb,amsfonts}
\usepackage{algorithmic}
\usepackage{graphicx}
\usepackage{textcomp}
\usepackage{xcolor}
\usepackage{float}
\usepackage{array}
\usepackage{booktabs}
\usepackage{amsmath}
\usepackage{multirow}
\usepackage{subcaption}

\usepackage{mathtools}
\usepackage{adjustbox}
\usepackage{tabulary}
\usepackage{colortbl}
\usepackage{array}
\usepackage[inkscapelatex=false]{svg}
\newcolumntype{H}{>{\setbox0=\hbox\bgroup}c<{\egroup}@{}}
\graphicspath{{./figures/}}

\def\BibTeX{{\rm B\kern-.05em{\sc i\kern-.025em b}\kern-.08em
 T\kern-.1667em\lower.7ex\hbox{E}\kern-.125emX}}
\begin{document}

\title{A PVT-Resilient Subthreshold SRAM-Based In-Memory Computing Accelerator with In-Situ Regulation for Energy-Efficient Spiking Neural Networks}
\author{\IEEEauthorblockN{
Shih-Hang Kao, Yang-Chan Hung, I-Wen Wang, Bing-Han Liu,Yu-Chia Chen, Tian-Sheuan Chang, \textit{Senior Member, IEEE},
Shyh-Jye Jou, Chien-Nan Liu, \textit{Senior Member, IEEE}, Hung-Ming Chen, \textit{Member, IEEE}, Wei-Zen Chen, \textit{Senior Member, IEEE}
}
\thanks{This work has been accepted to be published in IEEE Transactions on Circuits and Systems I: Regular Papers. Digital Object Identifier: 10.1109/TCSI.2026.3685722. This work was supported by the National Science and Technology Council, Taiwan, under Grant 113-2218-E-A49-024-MBK. The authors are affiliated with the Institute of Electronics, National Yang Ming Chiao Tung University, Taiwan. (e-mail:\{k880531.ee10, albert588.ee11, dylanr.ee10, binghan.liu20.11,neilchen065.ii12, tschang, jerryjou, jimmyliu, hmchen, wzchen\}@nycu.edu.tw }%
\thanks{Manuscript received XXXX XX, 2025; revised XXXX XX, XXXX.}
}

\maketitle

\begin{abstract}

This paper presents a PVT-resilient, subthreshold SRAM-based computing-in-memory (CIM) macro tailored for energy-efficient spiking neural networks (SNNs). The macro integrates in-situ current sensors and distributed voltage regulators to enable robust large-scale (1024 wordlines, 1304 bitlines and 128 shared neuron cells) subthreshold current-mode CIM, mitigating energy overheads and process-voltage-temperature (PVT) sensitivity. The neuron cells adopt a programmable, memory cell-based firing threshold to enhance neuron robustness against PVT variations. The architecture uses a stride-tick batching schedule to significantly reduce buffer overhead with enhanced input data reuse. Exploiting the high sparsity of SNNs, the proposed system demonstrates significant improvements in energy efficiency and variation tolerance. Fabricated in 28-nm CMOS, the prototype attains 93.64\% accuracy on keyword spotting, delivers up to 1181.42 TOPS/W, and achieves 7.24 TOPS/mm², demonstrating a viable and efficient solution for high-performance edge SNN processing.
\end{abstract}


\section{Introduction}

\begin{figure}[htbp]
\centering
\includegraphics[width=\columnwidth]{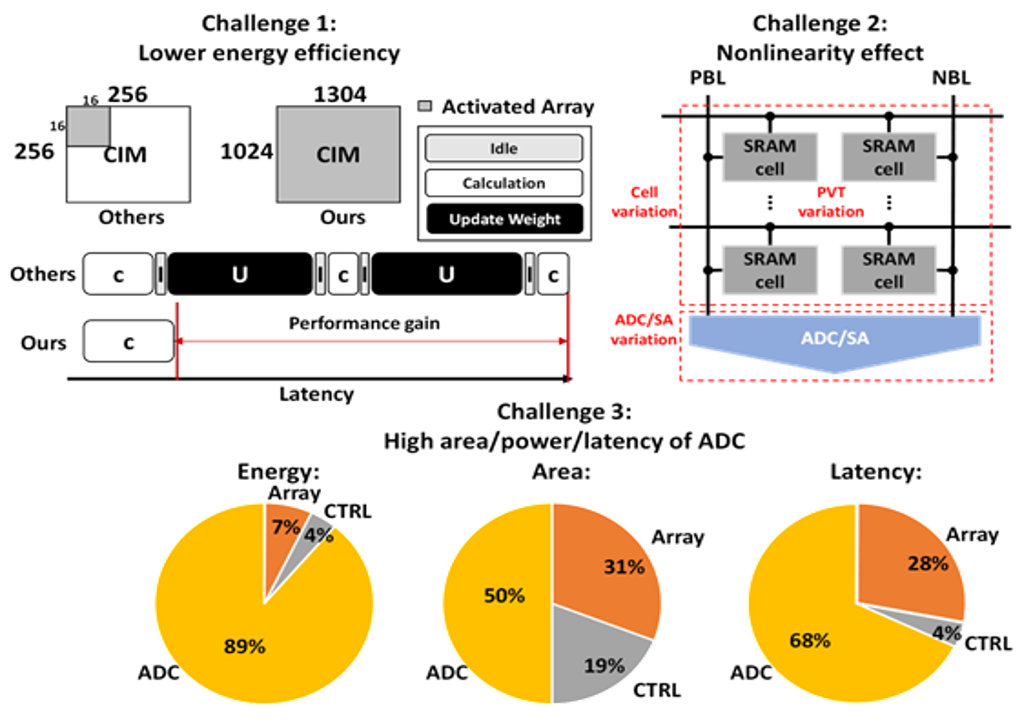}
\caption{The challenges of current CIM designs.}
\label{fig:challenges}
\end{figure}

Computing-in-memory (CIM) enables massively parallel computation with reduced data movement, making it an attractive architecture for AI acceleration. Nevertheless, conventional CIM realizations face three fundamental challenges (Fig.~\ref{fig:challenges}): (i) suboptimal energy efficiency due to small array size, (ii) accuracy degradation from circuit-level nonlinearities, and (iii) substantial power and area overheads from the analog-to-digital converter (ADC) arrays required for readout.

Spiking neural networks (SNNs) are a natural fit for CIM-based accelerators, exploiting event-driven computation and temporal sparsity. By substituting multi-bit ADCs with simple sense amplifiers (SAs), SNN-CIM systems can reduce power and improve noise tolerance. However, scaling SNN inference on CIM hardware introduces additional challenges, notably current-mode instability and pronounced sensitivity to process–voltage–temperature (PVT) variations. Prior SNN-CIM prototypes span RRAM~\cite{Tian2022} and eDRAM~\cite{Kim2023} technologies as well as SRAM-based designs~\cite{Liu2022, Kim2023b, Liu2024}. While these efforts pursue high energy efficiency and ADC-less operation~\cite{Kim2023b}, achieving robust, large-scale subthreshold operation with concurrent high accuracy and leading TOPS/W—particularly under PVT variation—remains difficult and is a central focus of this work.
Besides, SNN execution needs multiple timesteps, which requires large buffers to store the intermediate data for successive timestep processing. \cite{narayanan2020spinalflow} uses tick-batching that executes all timesteps first before processing to the next layer. But that is for digital design. \cite{SpikeSim} uses serial timestep and pipelined layer processing for CIM, which consumes significant buffer cost.

Traditional CIM multiply–accumulate (MAC) operations are implemented in either current mode or charge mode. Current-mode MACs \cite{Yu2022,Qiao2022} suffer from large output currents and limited bitline control, leading to power loss and readout saturation that impede large-scale parallelism. Charge-mode MACs \cite{Zang2023,Wang2023,Su2023} normalize the output dynamic range via charge sharing, alleviating overload concerns \cite{Jiang2020}; however, they require additional on-cell capacitors, making the MAC output sensitive to bitline parasitics and prone to nonlinearity in large arrays. Moreover, switch charge injection further degrades accuracy.

To address these challenges, this work propose a PVT-resilient, subthreshold SRAM-based CIM macro tailored for SNNs. At the circuit level, on-chip in-situ monitors and distributed regulators enable stable, low-power, large-scale current-mode operation with up to 1,024 simultaneously activated wordlines and 1,304 bitlines in the subthreshold regime— to the best of our knowledge, for the first time. The circuit introduces programmable, memory-cell–based neuron thresholds to improve robustness under PVT variation. The architecture employs a stride–tick batching schedule to enable higher input reuse and eliminate large intermediate buffers during multi-timestep execution. In addition, variation-aware training further hardens the model against SRAM device variation and sense-amplifier mismatch.

The remainder of this paper is structured as follows. Section~\ref{sec:circuit} presents the proposed circuit design. Section~\ref{sec:models} introduces the CIM-optimized SNN model, architecture, and dataflow. Section~\ref{sec:results} discusses experimental validation, and Section~\ref{sec:conclusion} concludes the paper.

\section{Circuit Design and PVT Compensation}
\label{sec:circuit}

\subsection{Subthreshold SRAM Operation}

\subsubsection{PVT Challenges in Current-Mode SRAM-CIM Macros}

\begin{figure}[tb]
\centering
\begin{subfigure}{\columnwidth}
\includegraphics[width=\textwidth]{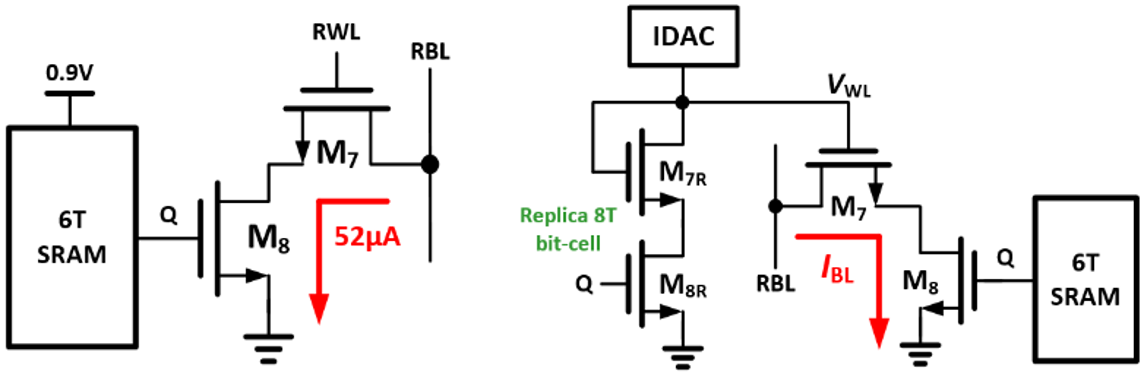}
\caption{}
\end{subfigure}
\begin{subfigure}{\columnwidth}
\includegraphics[width=\textwidth]{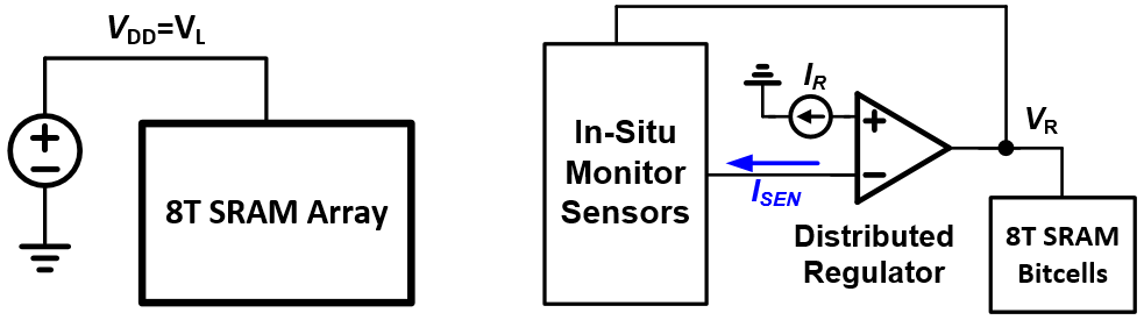}
\caption{}
\end{subfigure}
\caption{(a) Left: 8T SRAM output current under nominal supply voltage (0.9\,V). Right: 8T SRAM driven by a low WL voltage. (b) Left: 8T SRAM driven by a constant low supply voltage ($V_L$). Right: 8T SRAM driven by the proposed regulator.}
\label{fig:sram_cell}
\end{figure}

To realize an area/energy-efficient SNN accelerator, a current-mode CIM macro is preferable as the integration function is able to be easily implemented without an additional transducer. SRAM macros based on 6T SRAM cells offer the highest computing density due to their small cell size. However, they are susceptible to read disturbances when multiple wordlines are activated simultaneously. A more effective solution is to use 8T SRAM cells, which addresses this issue by incorporating additional coupling transistors ($M_7, M_8$) to separate read and write operations, as illustrated in left part of  Fig.~\ref{fig:sram_cell}(a). Despite this, it still faces significant challenges with the signal dynamic range issue in large-scale matrix operations. For example, the nominal readout current in an 8T SRAM cell under 0.9\,V supply is around 52\,$\mu$A. It is sensitive to supply voltage variations, leading to significant power consumption during large-scale current-mode operations. Additionally, it restricts kernel size to avoid overloading the MAC output. Hybrid architectures that combine analog CIM macros with digital near-memory post-processing units offer some relief \cite{Wu2023,Guo2024}, but still face issues with PVT sensitivity of analog partial sums before they are properly digitized. To reduce the output current of an 8T-SRAM cell, \cite{Zhang2016} lowers the wordline driving voltage (RWL) to drive transistor $M_8$ into the deep triode region while maintaining a nominal VDD voltage, as illustrated in right part of Fig.~\ref{fig:sram_cell}(a). This approach necessitates a current digital-to-analog converter (IDAC) to generate an appropriate WL voltage and ensure a constant bit cell output current. However, due to the dependence of the WL voltage response time on the unit cell charging current and WL parasitic capacitance, this method faces significant speed limitations in large-scale MAC operations. Therefore, we proposed an architecture that regulate the supply voltage ($V_L$) of the SRAM cell by a regulator to reduce the power consumption as illustrated in Fig.~\ref{fig:sram_cell}(b).

Existing CIM designs mainly address PVT-induced nonidealities through digital post-processing calibration~\cite{Bai2024}, or using replica-based reference array~\cite{Lee2024} for the readout circuit. In ~\cite{Bai2024}, compilation-aware frameworks are proposed to model nonidealities of CIM macro, such as ADC quantization noise and other voltage-induced errors, which are subsequently calibrated using weight-mapping strategies. Such methods can effectively recover algorithm-level accuracy and support flexible bit-width operations, but require additional calibration procedures that defer the latency of signal processing. On the other hand, replica columns or reference arrays~\cite{Lee2024} themselves still suffer from serious drifts problems over temperature variations, making it extremely challenging for a lower power and reliable operation under limited ADC dynamic range without adequate voltage regulation.

    In contrast, the proposed method directly regulates the computation current in the analog domain during operation, suppressing PVT-induced variations at their source, as is shown on the right of Fig. 2(b). Unlike calibration- or replica-based schemes, the regulation operates continuously without explicit calibration procedures or dedicated reference columns. Since on-chip voltage regulation (e.g., LDOs) is commonly required in a practical SoC system to ensure supply stability, the proposed regulation can be naturally integrated with existing analog biasing and regulation circuits. Specifically, the area overhead is less than  1.87$\%$.

\subsection{PVT Monitor Sensors and Distributed Regulators}
\label{sec:PVT_monitor}
\begin{figure}[tbp]
\centering
\includegraphics[width=\columnwidth]{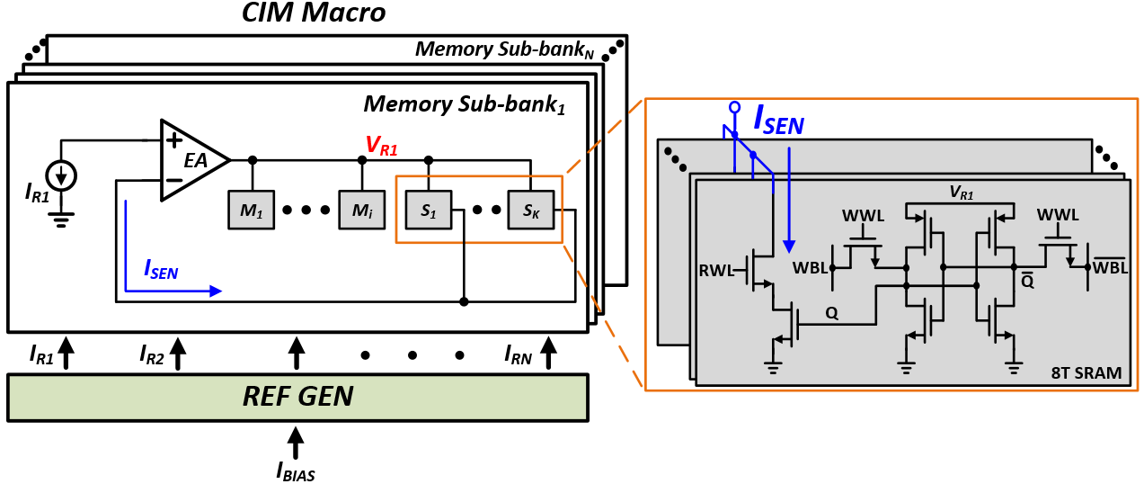}
\caption{CIM macro with monitor sensors and distributed regulators.}
\label{fig:cim_architecture}
\end{figure}

Fig.~\ref{fig:cim_architecture} depicts the circuit architecture of the CIM macro, which is segmented into several subbanks. Each subbank is powered by different supply voltages ($V_R$) through distributed regulators. Within each subbank, 10 out of 1024 SRAM cells ($S_K$, $K=1-10$) store a logic '1' and function as monitor sensors. With the wordline activated, in-situ sensing currents ($I_{SEN}$) collected from the monitor sensors are compared to reference current $I_{R1}$, which is derived from a global bias current ($I_{BIAS}$) via the reference generator (REF GEN). These currents are fed into distributed transimpedance error amplifiers (EA) that generate feedback voltages ($V_{R1}$) to individually power memory units ($M_1-M_i$) and monitor sensors ($S_1-S_{K}$) in each subbank, improving local and global current matching of the massive bit cells. As the unit current is normalized to global reference current, it prevents current overload across PVT variations in contrast to the prior art~\cite{Bai2024} ~\cite{Lee2024}.

\begin{figure}[tbp]
\centering
\includegraphics[width=\columnwidth]{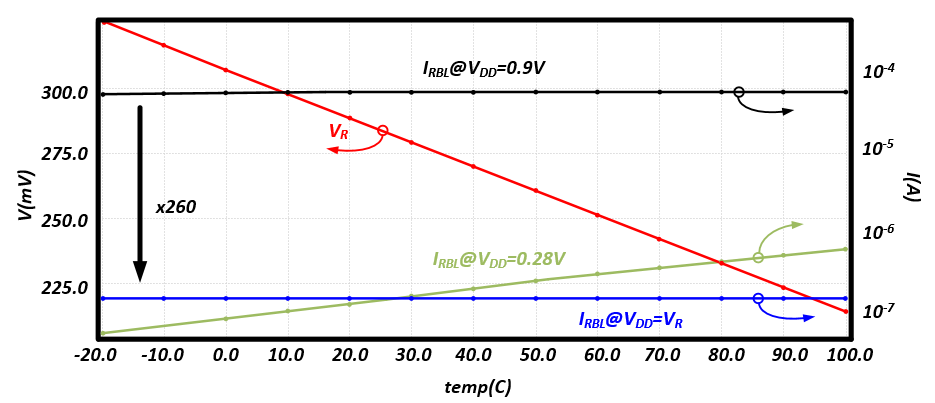}
\caption{Simulated RBL current variation at different temperatures. The arrows on the plotted lines are for y-axis reference.}
\label{fig:rbl_variation}
\end{figure}

Fig.~\ref{fig:rbl_variation} demonstrates the simulated bitline current drifts with/without the proposed scheme. Through the proposed technique, the supply voltage ($V_R$) of the SRAM cell is self-regulated from 219\,mV to 330\,mV across a temperature range of -20 to 100~$\mathrm{^\circ C}$, while consistently delivering a steady output current of 200\,nA. Compared to an 8T-bit cell operating under nominal 0.9\,V supply ($I_{RBL}$ @ $V_{DD}=0.9\,V$), the bit-cell output current ($I_{RBL}$ @ $V_{DD}=V_R$) is reduced by 260X, extending the dynamic range for dot-product operations by the same factor. Meanwhile, the static leakage current of the memory array is reduced by 87\%, decreasing from 385.86 nA to 48.99 nA, thanks to the reduced supply voltage. In contrast to an SRAM array powered by a voltage regulator, the CIM macro’s output current with a constant low supply voltage ($I_{RBL}$ @ $V_{DD}=0.29\,V$) varies by 8 times, impeding reliable operation of the sense amplifier or readout circuit. By properly programming the bit-cell current through $I_{BIAS}$, larger matrix operations become feasible without signal overloading issues.

\begin{figure}[htbp]
 \centering
 \begin{subfigure}{0.7\columnwidth}
 \includegraphics[width=\textwidth]{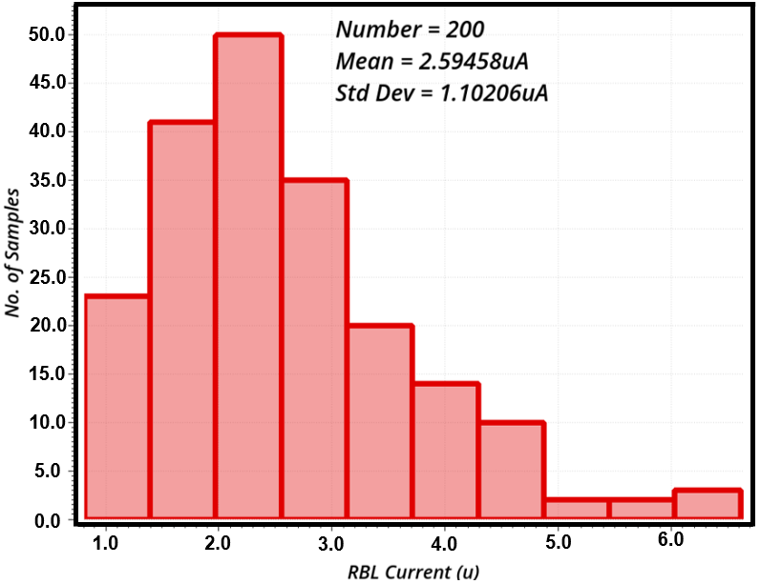}
 \caption{}
 \end{subfigure}
 \begin{subfigure}{0.7\columnwidth}
 \includegraphics[width=\textwidth]{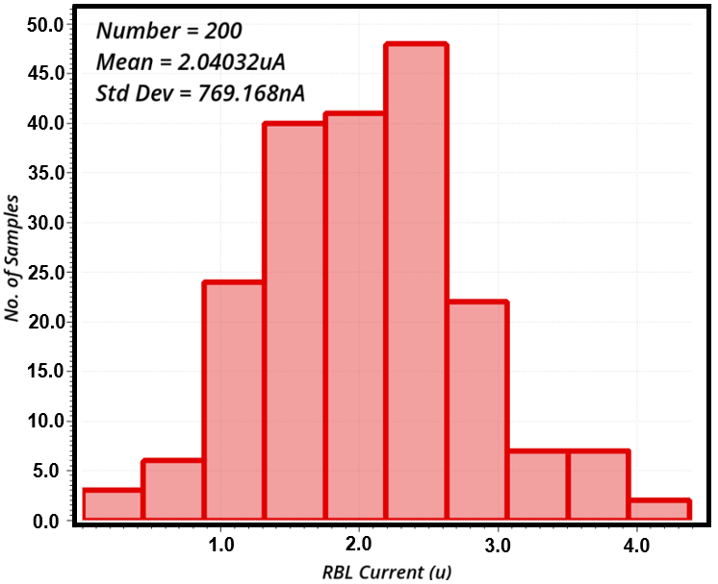}
 \caption{}
 \end{subfigure}
 \caption{Current variation of 8T SRAM cell (a) IDAC driven (b) proposed scheme.}
 \label{fig:current_variation}
 \end{figure}

Fig.~\ref{fig:current_variation}(a) and Fig.~\ref{fig:current_variation}(b) present the Monte Carlo simulations of bitline current variations (RBL) of 10 cells, comparing the IDAC-driven scheme shown on the right of Fig.~\ref{fig:sram_cell}(a) \cite{Zhang2016} to the  proposed bias scheme shown on the right of Fig.~\ref{fig:sram_cell}(b). The nominal unit current is set as 200\,nA in both cases. Results indicate that the mean and standard deviation of 10 cells output current are improved by 27.5\% and 43\%, respectively, with the proposed scheme at room temperature. In the IDAC-driven scheme (Fig.~\ref{fig:sram_cell}(a)), the read wordline (RWL) voltage is lowered to 418\,mV, leading to substantial $V_{DS}$ mismatches between transistors $M_8$ and $M_{8R}$ due to the channel length modulation effect in $M_7$ and $M_{7R}$. This mismatch degrades current matching between the replica and unit cell, as both operate in the deep triode region. The proposed scheme addresses this issue by setting the read wordline (RWL) to a nominal supply voltage of 0.9\,V, where $V_{DS}$ fluctuations at $M_8$ are negligible.

\begin{figure}[tbp]
\centering
\includegraphics[width=\columnwidth]{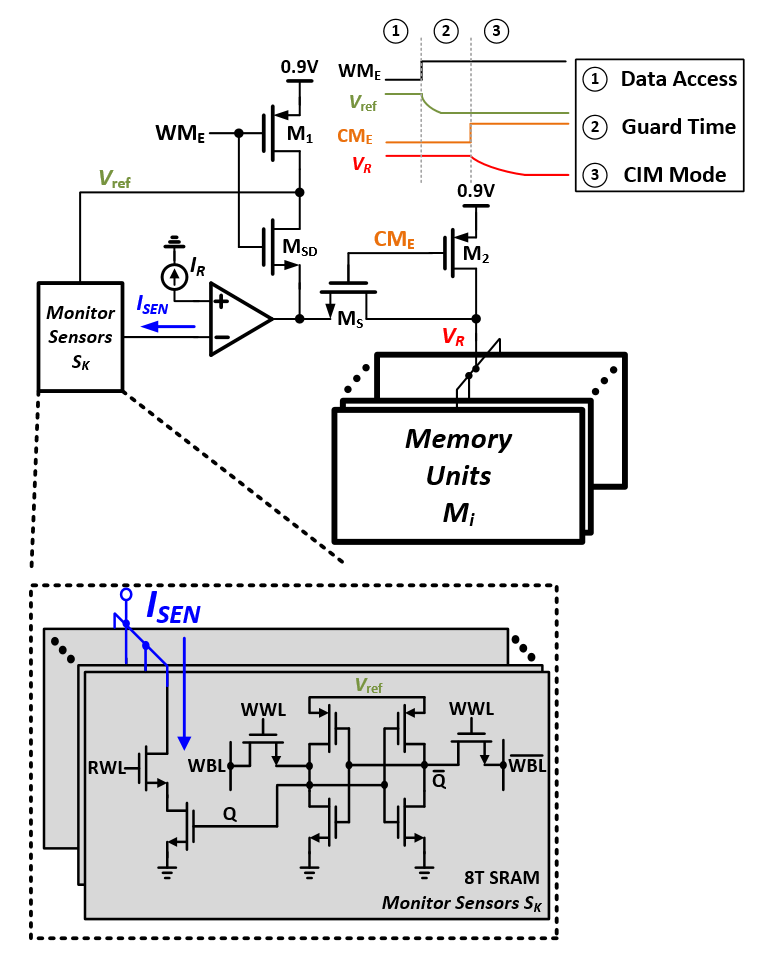}
\caption{The proposed voltage regulator and its timing diagram.}
\label{fig:voltage_regulator}
\end{figure}

The SRAM-CIM macro can operate in either data access mode, used for weight updates, or CIM mode, used for MAC operations. Fig.~\ref{fig:voltage_regulator} illustrates the supply switch configuration for each mode. 

In data access mode, both control signals—write mode enable ($WM_E$) and CIM mode enable ($CM_E$)—are low, allowing transistors $M_1$ and $M_2$ to turn on. This setup enables the SRAM subbank to operate at a nominal 0.9\,V supply, ensuring a fast response time and a high noise margin.

\begin{figure}[tbp]
\centering
\begin{subfigure}{0.6\columnwidth}
\includegraphics[width=\textwidth]{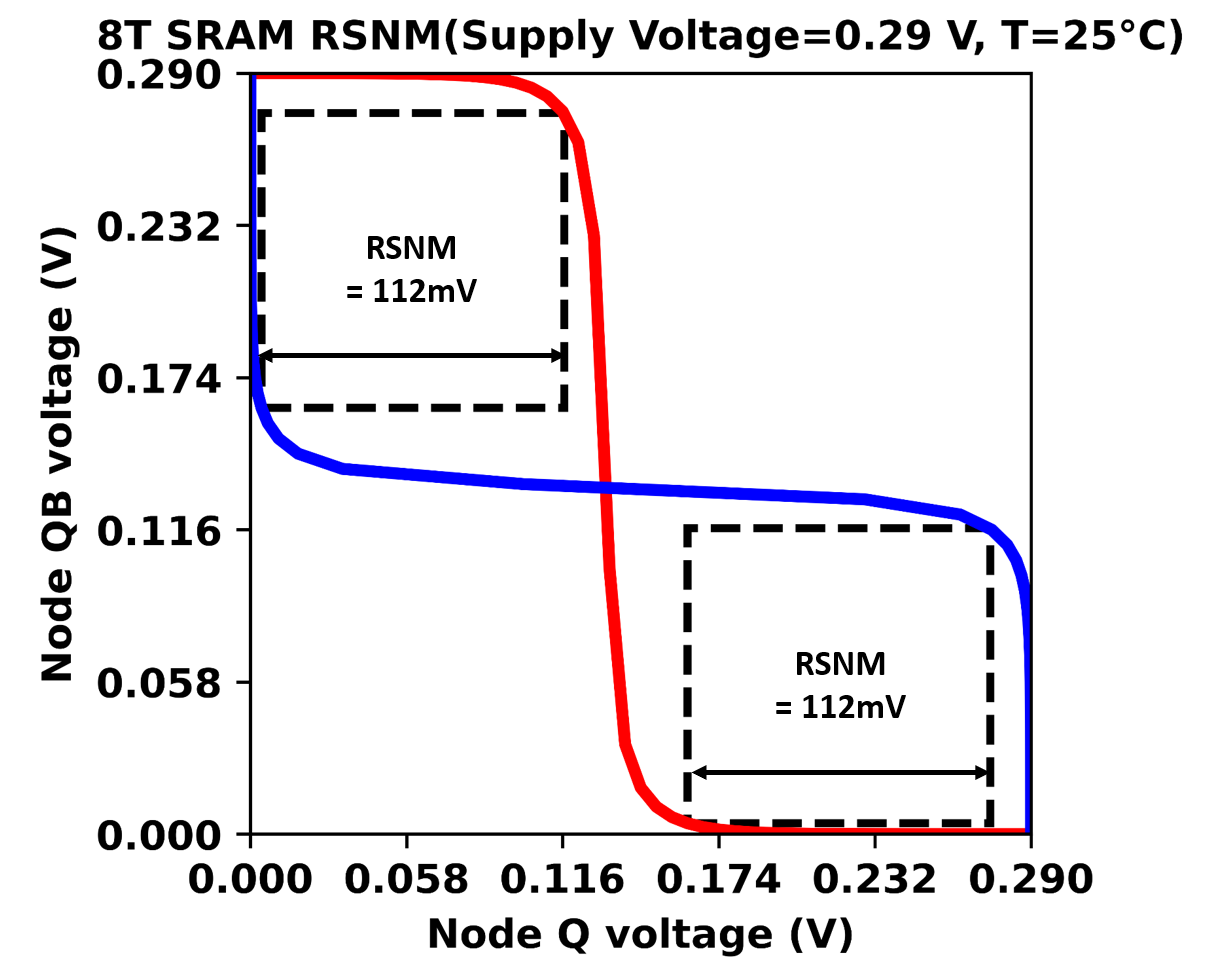}
\caption{}
\end{subfigure}
\begin{subfigure}{0.6\columnwidth}
\includegraphics[width=\textwidth]{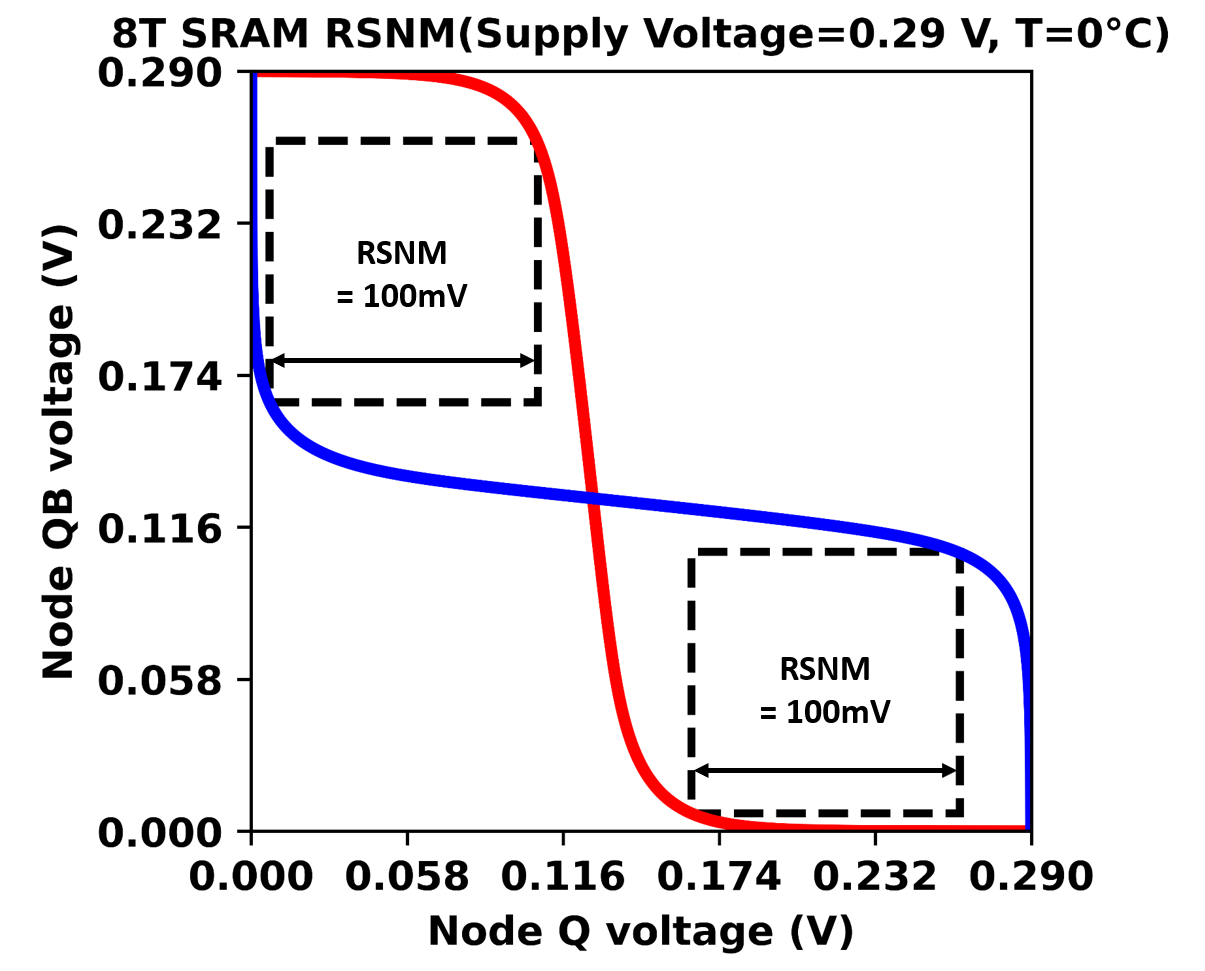}
\caption{}
\end{subfigure}
\begin{subfigure}{0.6\columnwidth}
\includegraphics[width=\textwidth]{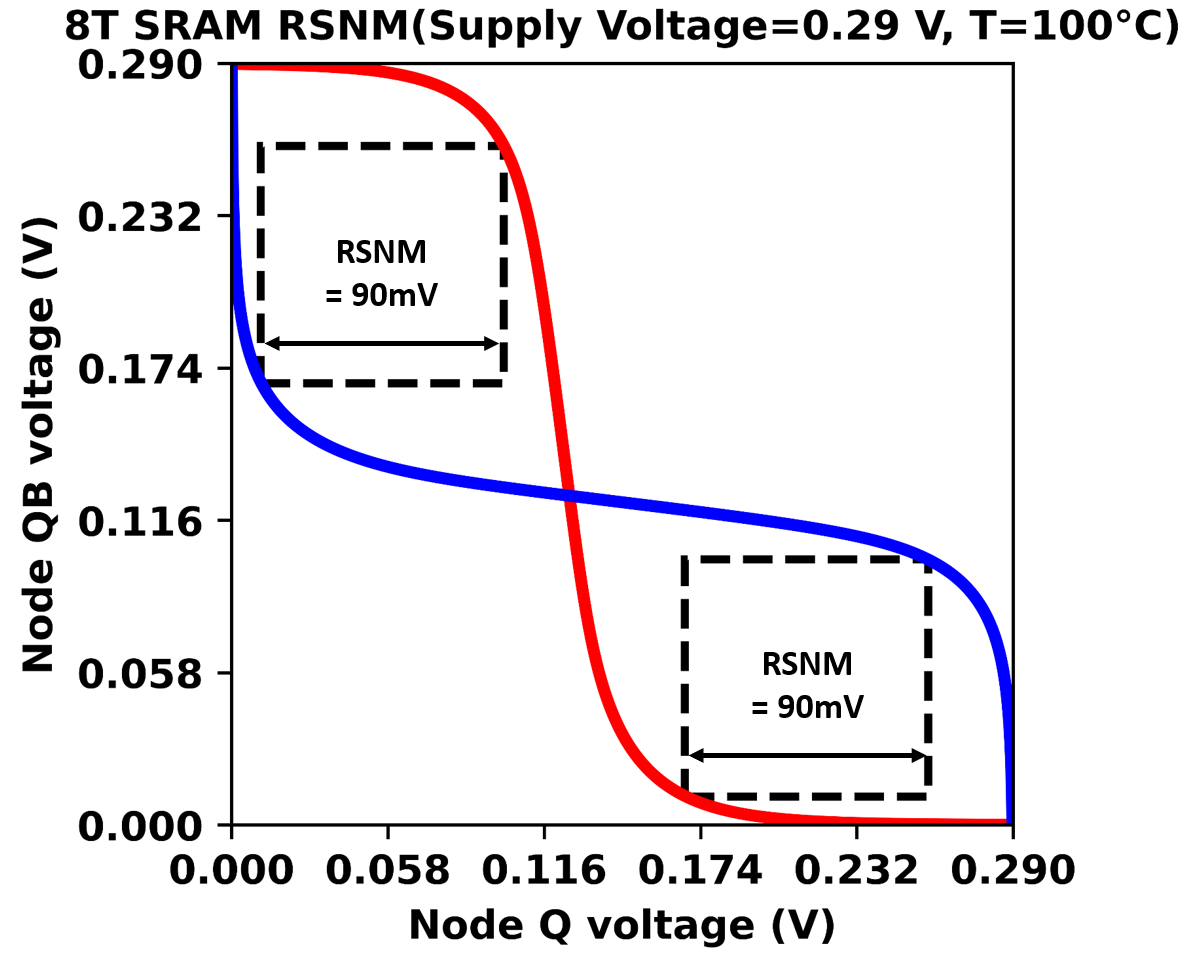}
\caption{}
\end{subfigure}
\caption{Read static noise margin of 8T SRAM (a)25℃ (b)0℃ (c)100℃.}
\label{fig:noise_margin}
\end{figure}

In CIM mode, $WM_E$ is set high to disconnect $M_1$ and activate $M_{SD}$,  while in-situ monitor sensors and distributed regulators are also engaged. The reference voltage ($V_{\text{ref}}$) then is reduced from 0.9 V by the distributed regulators, driving the monitoring devices into the subthreshold region, which in turn reduces output current. The control signal $CM_E$ is subsequently pulled high to disable $M_2$ and turn-on $M_5$. The regulated voltage ($V_R$) is then  applied to the memory subbanks. A guard time between activating $WM_E$ and $CM_E$ ensures $V_{\text{ref}}$ stabilization before it is applied to the memory bank, thereby minimizing power supply disturbance. The read static noise margin (RSNM) of the 8T SRAM in CIM mode is shown in Fig.~\ref{fig:noise_margin}. The RSNM remains at 112\,mV, even with the supply voltage reduced to 0.29\,V.

\begin{figure}[htbp]
\centering
\includegraphics[width=\columnwidth]{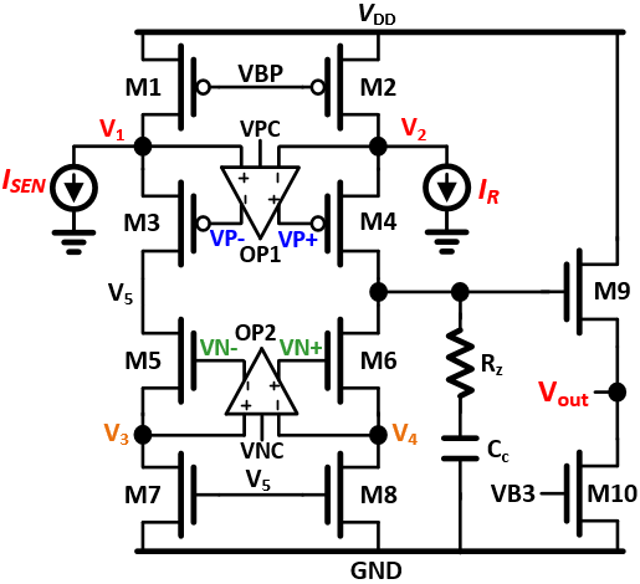}
\caption{Transimpedance error amplifier. }
\label{fig:amplifiers}
\end{figure}

Fig.~\ref{fig:amplifiers}(a) demonstrates the transimpedance error amplifier (EA) for the voltage regulator, in which $I_R$ is the reference current and $I_{SEN}$ represents the output current from the in-situ monitor sensors. The transimpedance gain is boosted by auxiliary amplifiers, OP1 and OP2. The current offset between $I_R$ and $I_{SEN}$ is reduced to 0.001\,\% through a high loop gain of 88\,dB.

\begin{figure}[htpb]
\centering
\begin{subfigure}{\columnwidth}
\includegraphics[width=\textwidth]{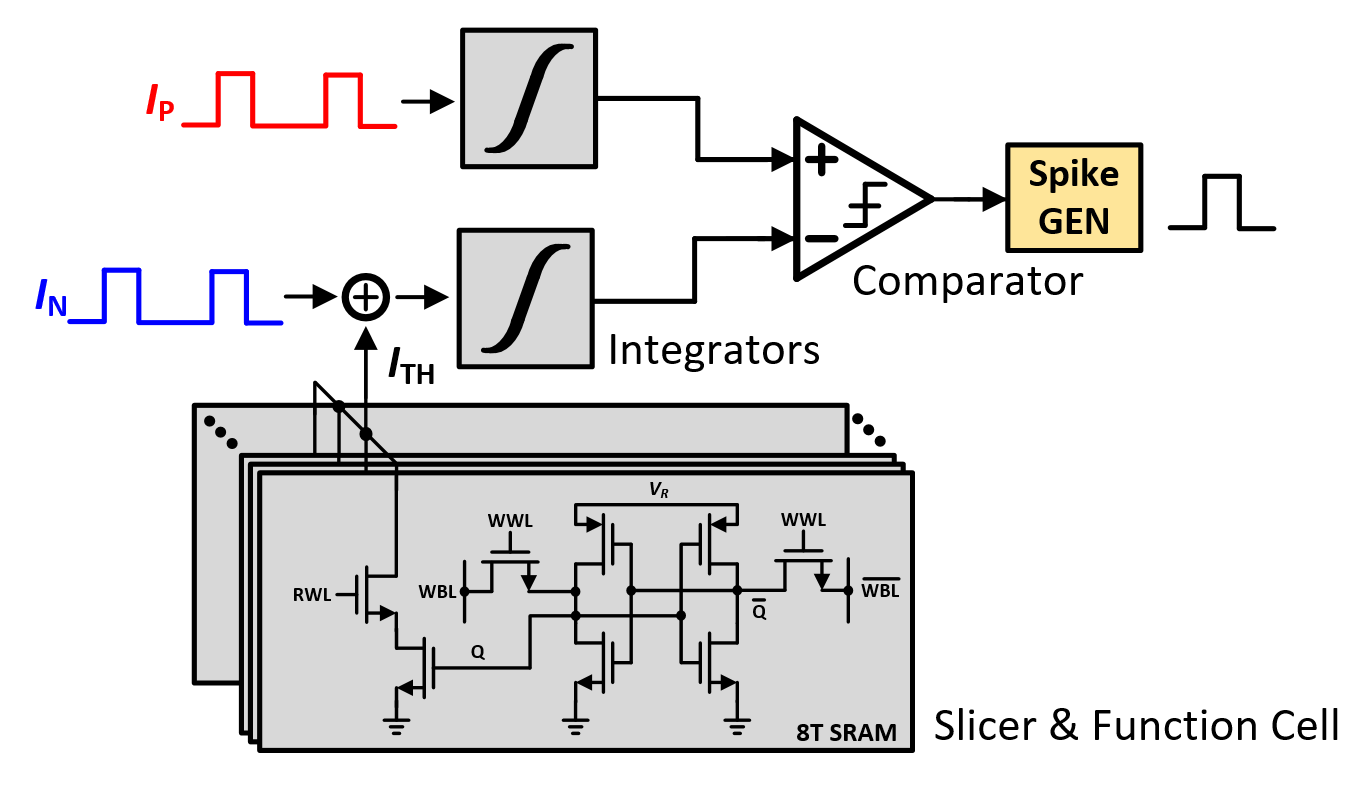}
\caption{}
\end{subfigure}
\begin{subfigure}{\columnwidth}
\includegraphics[width=\textwidth]{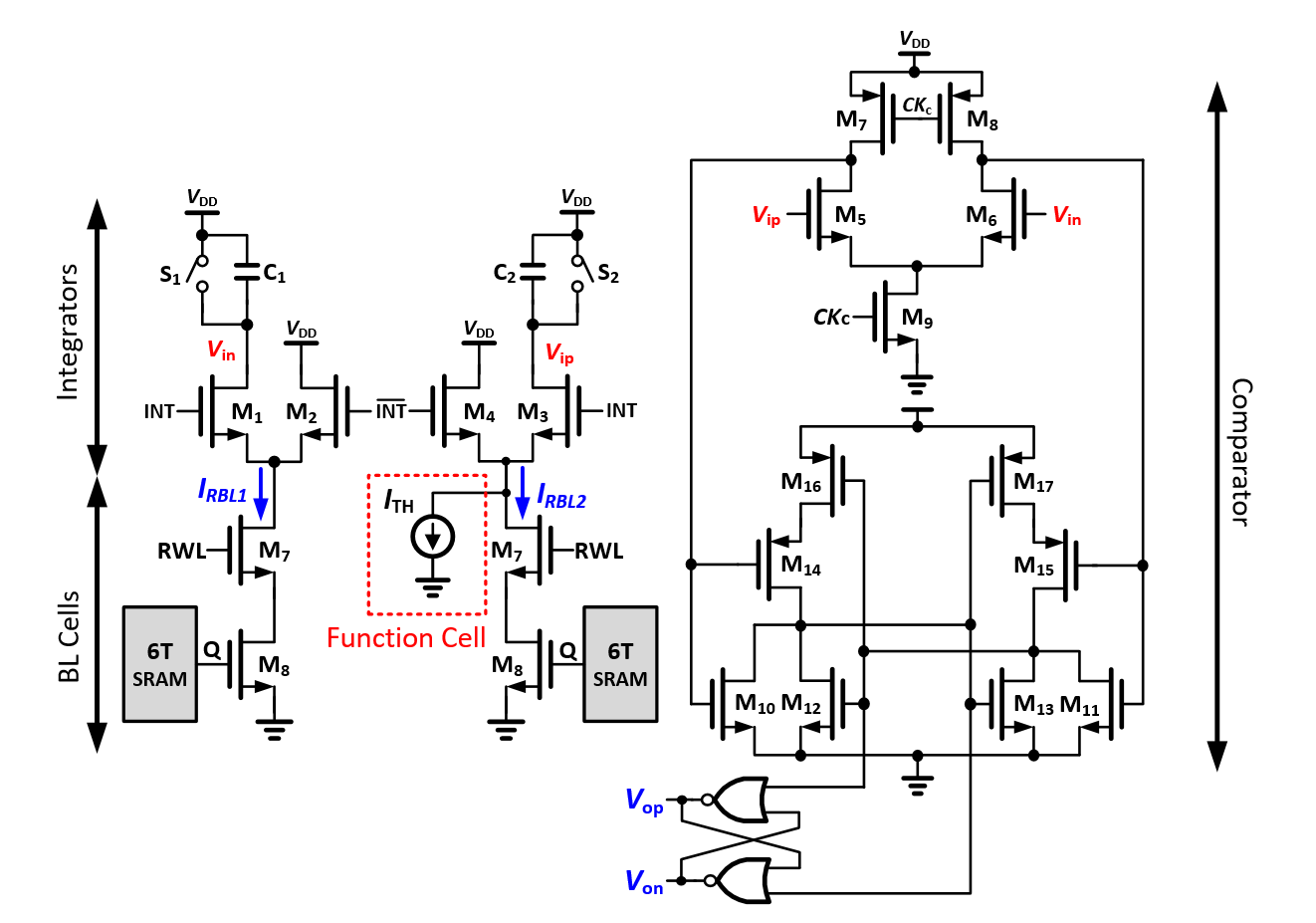}
\caption{}
\end{subfigure}
\begin{subfigure}{\columnwidth}
\includegraphics[width=\textwidth]{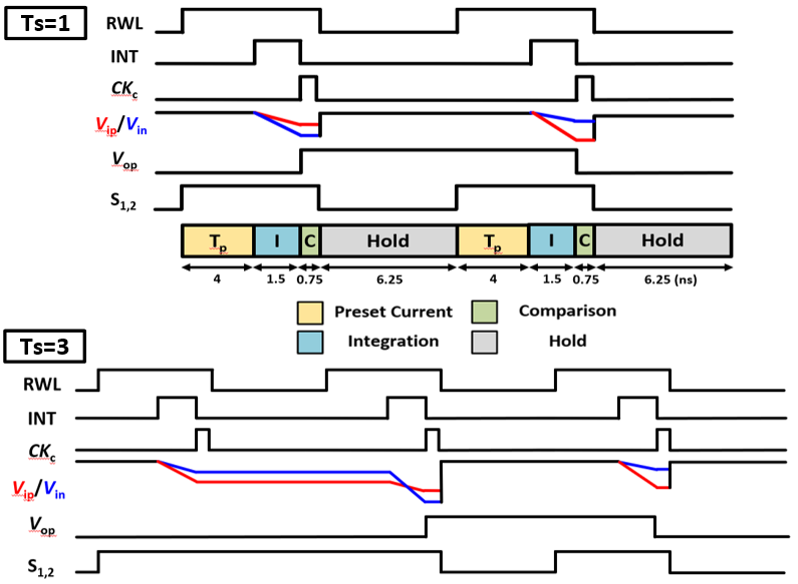}
\caption{}
\end{subfigure}
\caption{(a) The spike generation architecture. (b) SNN neuron. (c) Timing diagram of timestep of 1 and 3}
\label{fig:spike_gen}
\end{figure}

\subsection{SNN Neuron}

The membrane potential ($V_{\mathrm{mem}}[t]$) of a ternary-weight SNN is able to expressed as:
\begin{equation}
\small
\begin{aligned}
V_{\mathrm{mem}}[t] &= V_{\mathrm{mem}}[t-1] \cdot \left(1 - S[t-1]\right)
 + \sum_{i=0}^{1023} W_i \cdot IN_i[t], \\[4pt]
& \quad IN_i \in \{0,1\}, \; W_i \in \{-1,0,1\}, \\[4pt]
S[t] &=
\begin{cases}
1, & V_{\mathrm{mem}}[t] \ge V_{SNN_{th}}, \\
0, & V_{\mathrm{mem}}[t] < V_{SNN_{th}}.
\end{cases}
\end{aligned}
\label{eq:snn_membrane}
\end{equation}
The spike signal $S[t]$ is generated when the membrane potential $V_{\mathrm{mem}}[t]$
exceeds the threshold $V_{SNN_{th}}$.

To realize the neural function, the neuron cell is implemented as shown in Fig.~\ref{fig:spike_gen}(a), which incorporates integrators, a slicer, and function cells (FC) that generate the threshold current. 
Fig.~\ref{fig:spike_gen}(b) demonstrates the detailed circuit schematic. By the nature of current mode operation, the ternary weight dot-product currents are integrated on capacitors $C_1$ and $C_2$. The integrated outputs $V_{ip}$ and $V_{in}$ are subsequently compared by the comparator.
Theoretically, the neuron cell generates a spike when the membrane potential (i. e., voltage difference between $C_1$ and $C_2$) exceeds the threshold $V_{SNN_{th}}$. As the voltage threshold is highly sensitive to parasitic effects and depends on the input offset voltage of the comparator, the FC offers a programmable firing threshold ($I_{TH}$) at the integrator input instead to mitigate this effect. The $I_{TH}$ is generated from replica SRAM cells to accommodate potential random current variations across the array. Compared to generating $V_{SNN_{th}}$, the proposed $I_{TH}$ scheme circumvents design challenges of implementing transimpedance gain that should be adjusted across  PVT variations. In this design, the $I_{TH}$ corresponds to the output current of five unity cells. The proposed circuit consumes only 0.9$\%$ of the total chip power (12.39 mW), where each sense amplifier and the $I_{TH}$ generator consume 25.2 $\mu$W and 0.9 $\mu$W, respectively. The total area overhead is 364 $\mu$$m^2$ for 128 instances, corresponding to only 0.0011$\%$ of the total chip area (3.28 $mm^2$). Overall, it allows for precise and programmable control of the threshold current, reducing the impact of parasitic capacitance on the threshold and thereby improving the system's stability and accuracy.

The CIM macro supports Timestep = 1 to Timestep = 3 operations. Fig.~\ref{fig:spike_gen}(c) illustrates the timing diagrams, which consists of preset, integration, comparison, and hold phases. For Timestep = 1 operation ($T_s$=1), the SNN neuron resembles a CNN neuron cell, in which $I_{TH}$ is injected before comparison to create a threshold for the data slicer. The SNN output is shown as $V_{op}$, and the membrane potential is reset every cycle through switches $S_1$ and $S_2$.  
Timestep = 2 and Timestep = 3 operate similarly. The timing diagram for Timestep = 3 operation (Ts=3) is shown in detail as an example. Here three clock cycles are processed as a single functional group. The threshold current $I_{TH}$ is applied during the first cycle and is  withheld until either a spike is triggered ($V_{op}$=1) or the 3 cycle-duration elapses. Once one of these conditions is met,  the integration and comparison processes resume.

\section{Models, System Architecture and Dataflow}
\label{sec:models}
\subsection{CIM-Friendly SNN Model}
To support energy-efficient in-memory computing, the proposed system adopts an SNN model designed to eliminate the need for multi-bit ADCs. SNN encodes input with multiple spikes over time, integrates these input spikes and fires a spike output if summation of the input accumulation and current membrane potential exceeds the potential. In contrast to conventional ANN-based CIM accelerators, which rely on complex readout circuits, the binary outputs of SNNs allow simpler and more energy-efficient SAs for spike detection. SNNs emit spikes only when membrane potentials exceed a threshold, resulting in sparse activity and reduced switching. This event-driven behavior makes them highly compatible with ultra-low-power hardware deployment. Additionally, single-bit SNN neurons offer greater robustness to circuit nonlinearities than multi-bit ADCs.

Despite this advantage, implementing SNNs in large-scale CIM systems introduces practical challenges. These include extended computation latency due to multi-time-step integration and the memory overhead of storing intermediate membrane potentials. Furthermore, typical normalization operations for SNNs, such as batch normalization, cannot be implemented within a CIM array, as they require multi-bit outputs and ADCs. These issues are addressed through model and architecture co-design in this work.

\begin{figure}[tbp]
\centering
\includegraphics[width=\columnwidth]{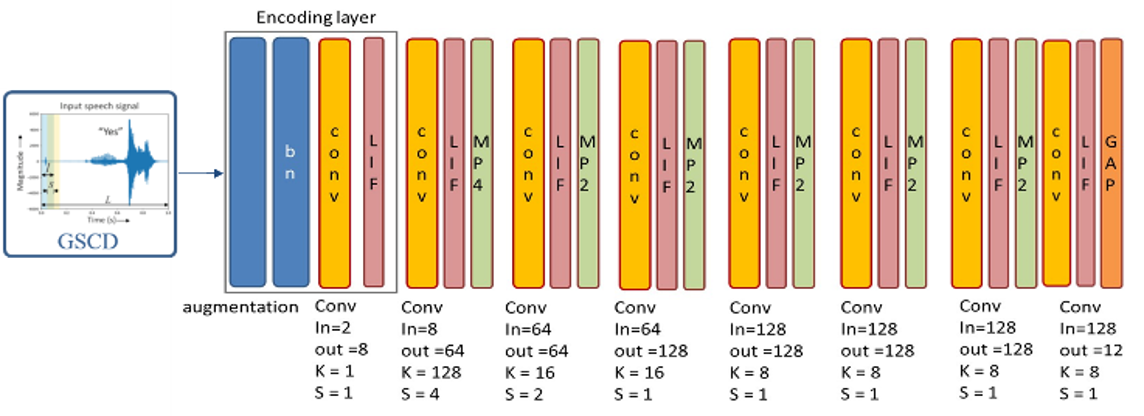}
\caption{The proposed example SNN model for keyword spotting.}
\label{fig:keyword_model}
\end{figure}

Fig.~\ref{fig:keyword_model} demonstrates the proposed hybrid SNN model targeted for keyword spotting as an example, where \textit{In/Out} denotes the number of input and output channels per layer, $K$ represents the kernel size ($K\times1$), and $S$ indicates the max pooling size ($S\times1$). \textit{Conv} refers to the convolution layers and max pooling layers, while \textit{LIF} stands for the leaky integrate-and-fire (LIF) neuron. In this model, only the input encoding layer includes a batch normalization layer, while the remaining seven convolution blocks are designed without normalization. These normalization-free blocks are able to be efficiently implemented on CIM hardware for computation without requiring high-precision ADCs. Each of the seven normalization-free blocks, except for the final block, consists of a convolution layer followed by a max pooling layer. In the final block, the LIF neuron is removed to allow the membrane potential to accumulate across all timesteps using an additional accumulator, which is then passed through an average pooling layer to serve as input to the classifier. In this design, the whole model except the input encoding and final output layers is implemented in this CIM.

To balance accuracy and efficiency, the model uses binary input activations and ternary weights, trained via spatio-temporal backpropagation with progressive quantization. The required number of timesteps is initially set to 3, and is able to reduced to 1 using the progressive timestep pruning method \cite{Chowdhury2021}, allowing for the selection of timestep numbers ranging from 1 to 3 during model inference. Meanwhile, fewer timesteps lead to reduced computation time on hardware and smaller feature map storage, creating a trade-off between performance and computation cost. Stride-tick batching is introduced to support efficient multi-timestep inference in hardware, minimizing buffer storage and processing delay.

\subsubsection{Variation-Aware Training}
\begin{figure}[htbp]
\centering
\includegraphics[width=\columnwidth]{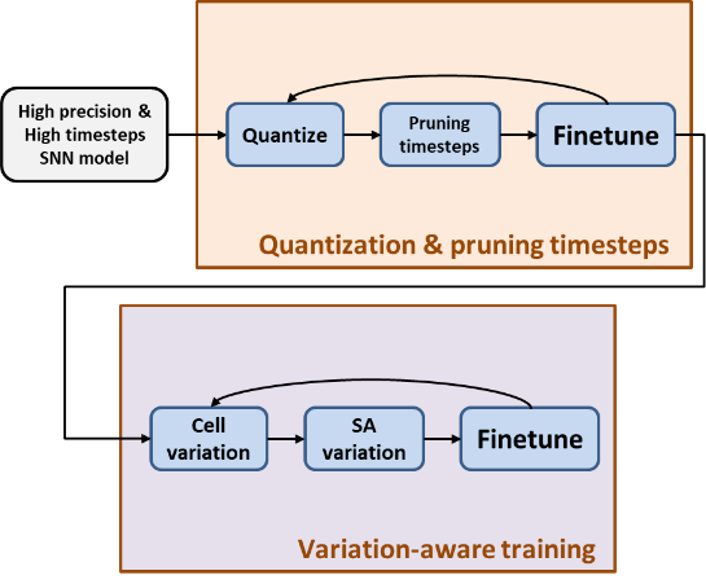}
\caption{SNN model training flow.}
\label{fig:training_flow}
\end{figure}

Hardware implementations of CIM inevitably introduce non-idealities such as input offsets in the SA, current mismatches in the SRAM macro, and noise contributions from the unit cell and readout circuit. To address these challenges, the SNN model is trained using a variation-aware refinement process, as depicted in Fig.~\ref{fig:training_flow}. The SNN is initially pretrained as a high-precision model operating over three timesteps. After quantization and timestep pruning, the model is further fine-tuned to account for the non-ideal effects of CIM. These variations are characterized based on Monte Carlo circuit simulations, and all extracted variations are incorporated into the variation-aware simulation.

According to Monte Carlo simulations, the input-referred offset and noise distributions at the SA input are approximately 7.28~$mV$ and 1~$mV_{rms}$, respectively. By considering these non-idealities under different temperature, Table~\ref{tab:variation_training} summarizes the performance comparison before and after variation-aware training. Using the ideal model's accuracy of 96.58\% as a benchmark, the accuracy degrades to 59.64\% without model adjustment but is able to recover to 93.64\% through the variation-aware refinement process.

\begin{table}[htbp]
\centering
\caption{Results of applying variations and variation-aware training on the proposed SNN model.}
\begin{tabular}{lc}
\toprule
\textbf{Condition} & \textbf{Accuracy (\%)} \\
\midrule
Ideal Model & 96.58 \\
With Variations (No Adjustment) & 59.64 \\
With Variation-Aware Training & 93.64 \\
\bottomrule
\end{tabular}
\label{tab:variation_training}
\end{table}

\begin{figure}[htbp]
\centering
\includegraphics[width=\columnwidth]{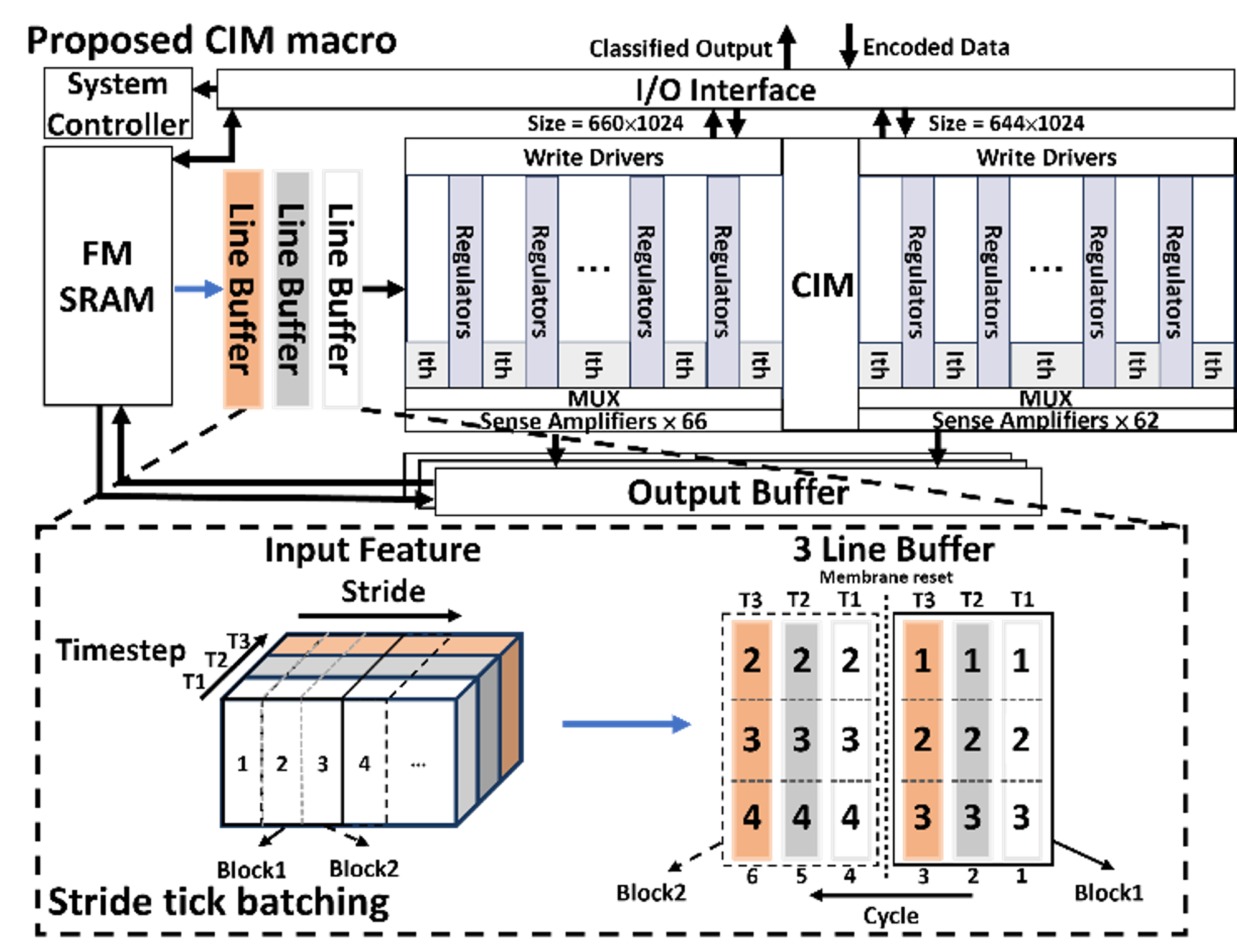}
\caption{Proposed CIM macro and stride-tick batching dataflow.}
\label{fig:cim_macro}
\end{figure}

\begin{figure}[htbp]
\centering
\includegraphics[width=\columnwidth]{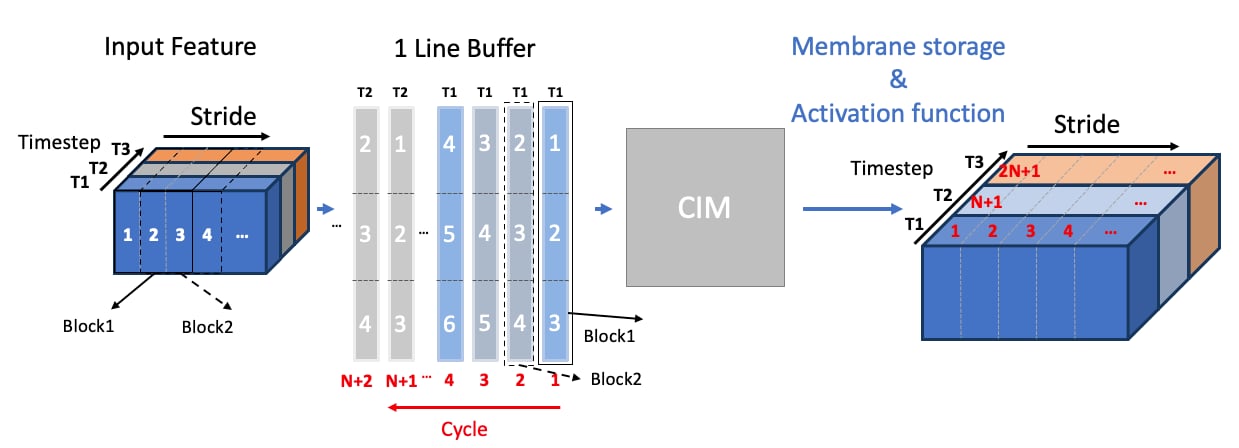}
\includegraphics[width=\columnwidth]{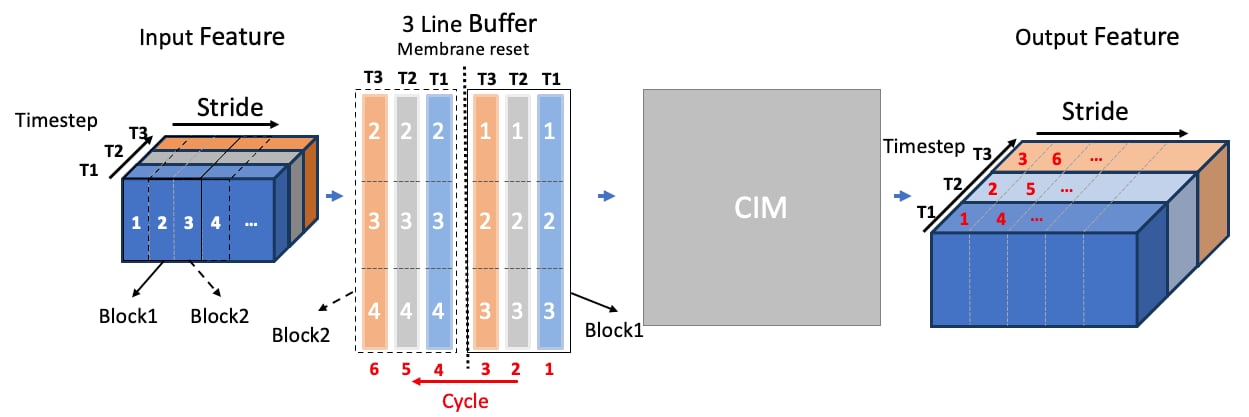}
\includegraphics[width=0.44\columnwidth]{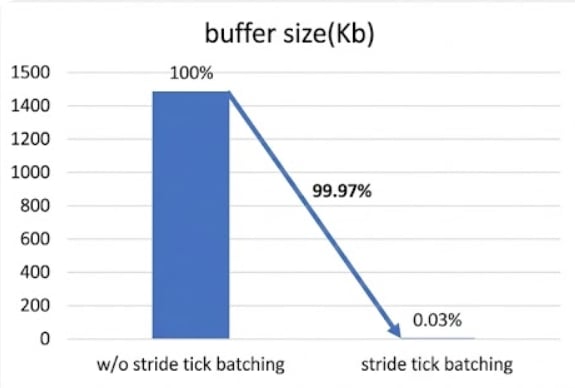}
\includegraphics[width=0.54\columnwidth]{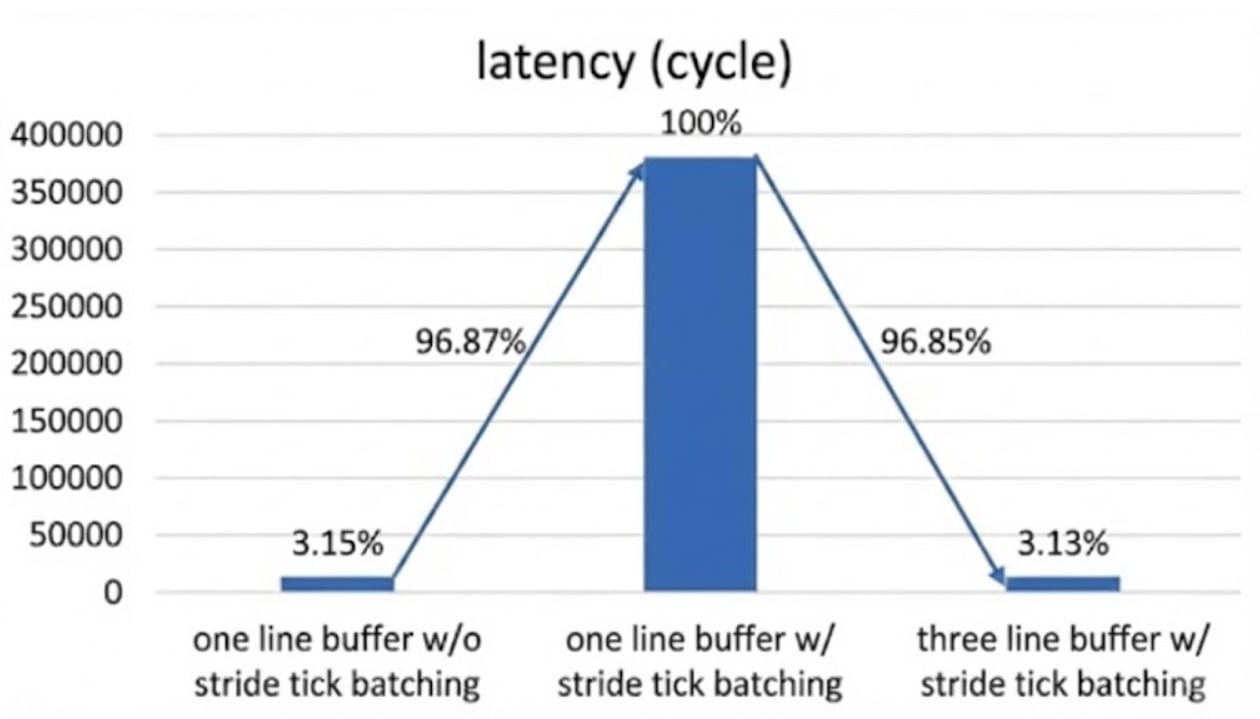}
\caption{Typical SNN and stride-tick batching dataflow.}
\label{fig:dataflow}
\end{figure}

\subsection{Hardware Architecture}
Fig.~\ref{fig:cim_macro} demonstrates the overall CIM macro architecture. The CIM macro integrates two subarrays to optimize layout aspect ratio and maximize parallelism. A full-row SRAM activation allows complete weight vector computation without partial sums, thereby removing the need for ADCs and improving energy efficiency. The two subarrays are operated in the subthreshold region for lower power. Thus, to ensure PVT-resilient subthreshold operation and high linearity at the MAC output, this design embeds in-situ current sensors and distributed voltage regulators within the array and leverages the high sparsity of the SNN to enable robust subthreshold current-mode computation. These circuit features ensure high linearity and resilience to PVT variation in large-scale inference workloads.

\subsubsection{Dataflow with Stride-Tick Batching}

Implementing SNNs in CIM architectures typically faces challenges with large buffer requirements for intermediate membrane potential storage, especially with multi-timestep operations. For instance, a conventional step-by-step flow (as conceptually illustrated for comparison in Fig.~\ref{fig:dataflow}) would require buffer space equivalent to the total feature map size, which is 1488~Kb in our model. Additionally, processing data step-by-step with a single line buffer can lead to significant latency in loading features for subsequent timesteps, as shown in the latency comparison in Fig.~\ref{fig:dataflow} (for the single line buffer without stride-tick batching case, though stride-tick itself also faces latency challenges if not optimized).

To address the storage issue, this paper proposes the stride-tick batching dataflow based on~\cite{narayanan2020spinalflow}, as shown in Fig.~\ref{fig:cim_macro}. In this flow, the CIM macro computes all timesteps (T1, T2, T3) for a given input feature block (e. g., block1) successively and jumps to the next one based on the stride setting. This allows for successive potential accumulation in the neuron for that block across its timesteps, and then the neuron is reset for the next input feature block (e. g., block2). This approach effectively eliminates the need for ADCs or large extra digital buffers to store intermediate membrane potentials between timesteps within a block's processing. Instead, the membrane potential is accumulated on a capacitor within the neuron cell before the sensing amplifier. As shown in the buffer size comparison in Fig.~\ref{fig:dataflow}, this reduces the digital equivalent storage for membrane potentials from 1488~Kb (for a step-by-step flow storing full feature maps) to just 0.375~Kb, achieving a 99.97\% reduction.

While stride-tick batching effectively reduces storage, a naive implementation with a single shared line buffer for input features across all timesteps would lead to a severe lack of feature reuse. This would significantly increase the input loading latency. For instance, as illustrated by the "one line buffer with stride tick batching" case in the latency graph of Fig.~\ref{fig:dataflow}, this could increase the latency for processing the first layer from 12,000 cycles (typical step-by-step with a single line buffer, no stride-tick) to 380,928 cycles.

To mitigate this latency increase while retaining the storage benefits of stride-tick batching, our proposed hardware architecture (Fig.~\ref{fig:cim_macro}) employs three separate line buffers, one dedicated to each potential timestep (T1, T2, T3). As input features for a block (e. g., Block1) are read from the FM SRAM, they are distributed to these respective line buffers. This allows input features for subsequent blocks (e. g., Block2) to be pre-fetched and stored while the current block is being processed across its timesteps, enabling independent processing of features for each timestep and significantly improving input data reuse. As seen in Fig.~\ref{fig:cim_macro}, for Block2, the features needed for T1, T2, and T3 are readily available from their respective line buffers. This strategy increases input data reuse from 0\% (with a single line buffer and stride-tick batching) up to 66\% (when considering the reuse for Block 2 as shown in Fig.~\ref{fig:cim_macro}) for a 3-timestep operation. Consequently, the latency is reduced back to 11,936 cycles for the first layer, as shown by the "three line buffer with stride tick batching" case in Fig.~\ref{fig:dataflow}'s latency graph. This design, therefore, balances storage efficiency, latency, and hardware resources, supporting programmable operations with 1 to 3 timesteps. This significant latency reduction comes with the manageable trade-off of incorporating two additional 1024-bitline buffers compared to a single line buffer setup, optimizing the overall dataflow for the SNN operations.

\subsubsection{Pipelined CIM Computation and Pooling}

\begin{figure}[htb]
\centering
\includegraphics[width=\columnwidth]{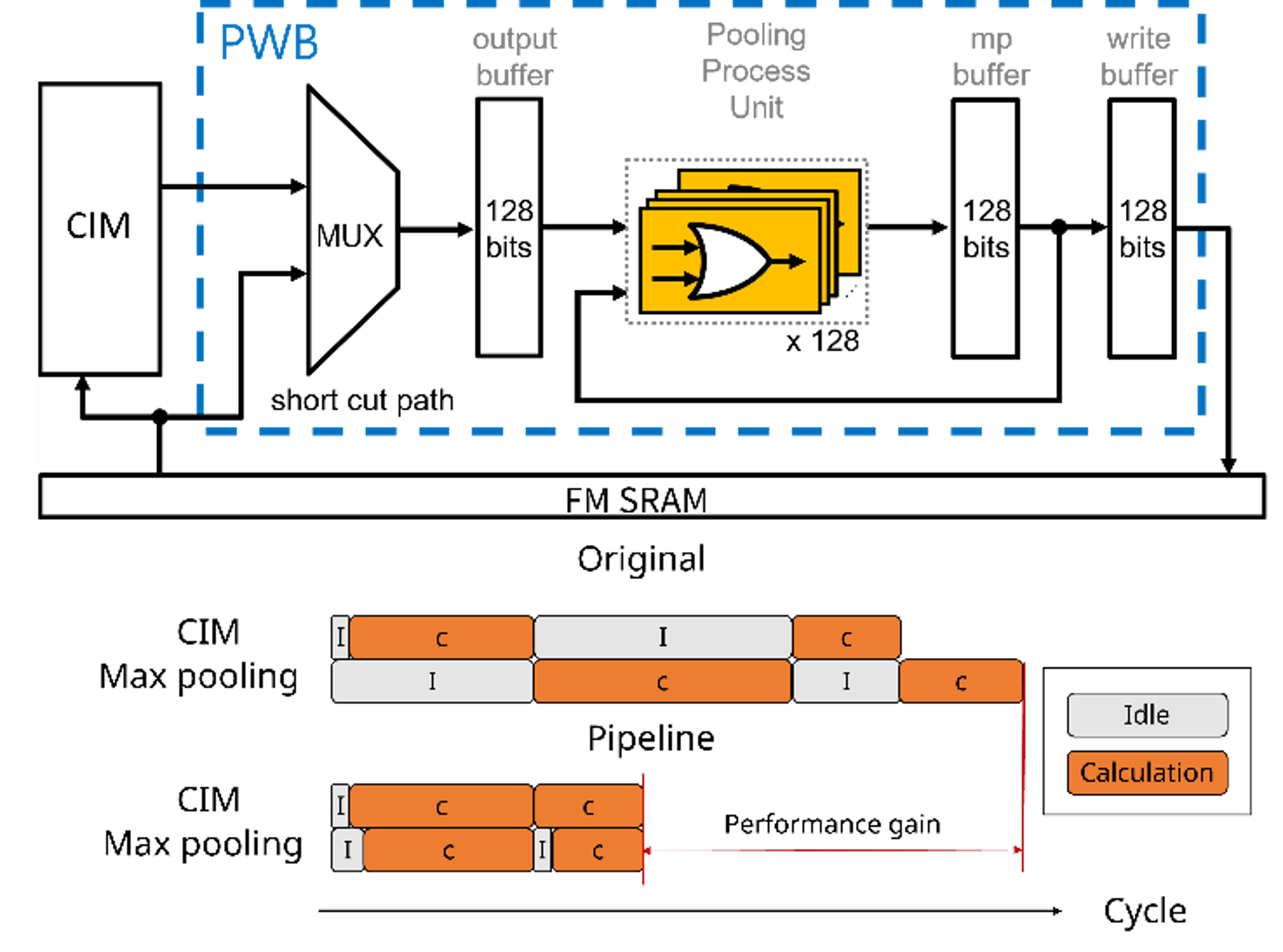}
\caption{Pooling write-back (PWB) timing diagram.}
\label{fig:pooling_write}
\end{figure}

Max pooling is widely utilized in models to extract important features from the feature map while reducing its size. However, when models are executed layer by layer, the CIM macro becomes idle during max pooling computation, leading to inefficiencies. To address this issue, pipelining the execution of max pooling and convolution operations is proposed as a solution. For this purpose, a pooling write-back (PWB) is implemented as shown in Fig.~\ref{fig:pooling_write}. Given that the activations are binary, the max pooling operation is able to be efficiently computed using OR gates. This approach allows for the simultaneous execution of convolution and max pooling operations, thereby reducing overall latency and improving computational efficiency.

With the implementation of the pooling write-back, the CIM computation and pooling are able to be fully pipelined. Using our keyword spotting model as an example, instead of performing convolution followed by pooling and then moving to the next convolution layer, the pipeline processing allows the next convolution layer to commence immediately after the previous one completes. This optimization reduces latency from the original 9873 cycles to 4945 cycles, achieving a 49.92\% reduction in latency.

\begin{figure}[htbp]
\centering
\includegraphics[width=\columnwidth]{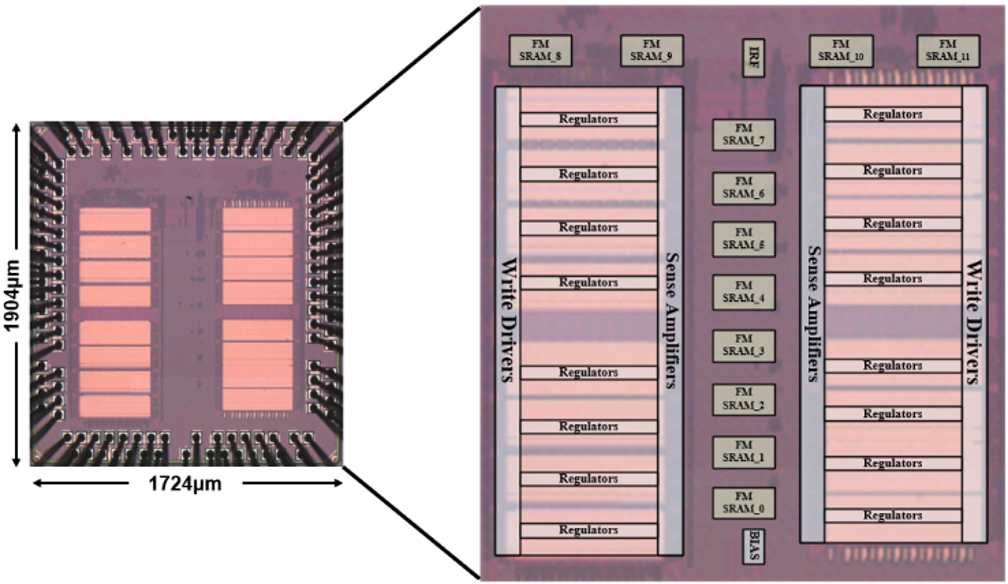}
\caption{Chip micrograph.}
\label{fig:chip_micrograph}
\end{figure}

\begin{figure}[htbp]
\centering
\includegraphics[width=\columnwidth]{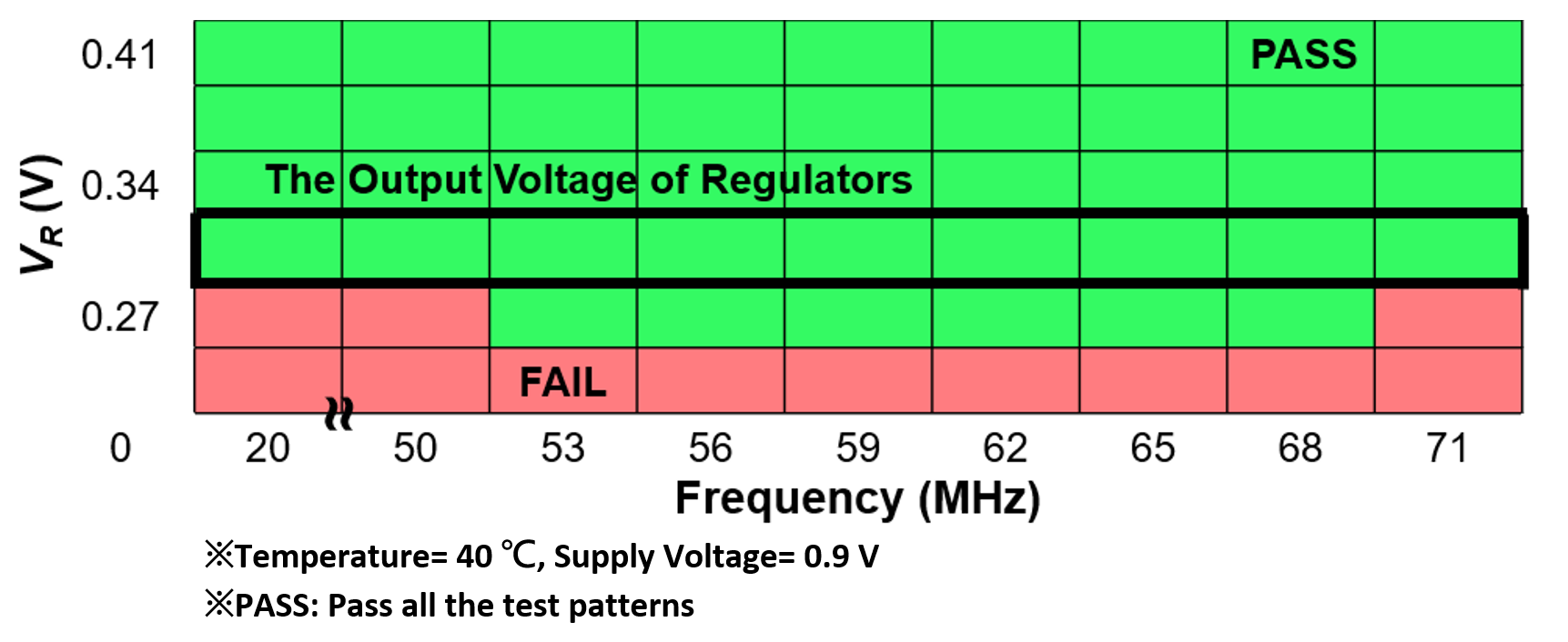}
\caption{Shmoo plot.}
\label{fig:shmoo_plot}
\end{figure}

\begin{figure}[htbp]
\centering
\includegraphics[width=0.8\columnwidth]{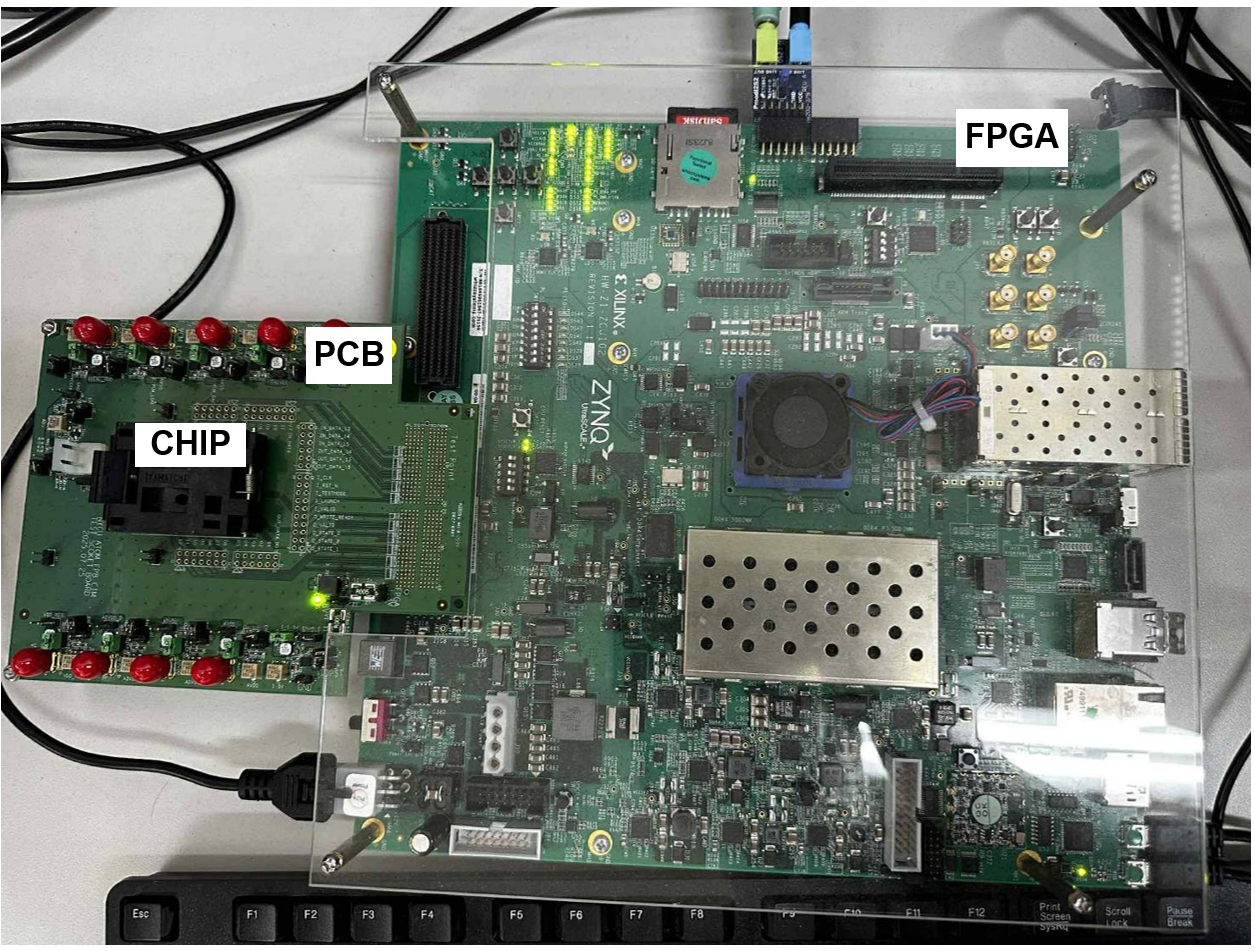}
\caption{Experimental setup.}
\label{fig:Experience_setup}
\end{figure}

\begin{table*}[htbp]
\centering
\caption{Comparison with previous designs}
\begin{tabular}{|c|c|c|c|c|c|c|}
\hline
 & ISSCC'22\cite{Liu2022} & ISCAS'22\cite{Tian2022} & ISCAS'23 \cite{Kim2023} & JSSC'23\cite{Kim2023b} & ISSCC'24 \cite{Liu2024} & This Work \\
\hline
Technology & 180 nm & 32 nm & 28 nm & 28 nm & 22 nm & 28 nm \\
\hline
Classifier & SNN & SNN & SNN & SNN & SNN & SNN \\
\hline
Area (mm$^2$) & 10 & 3.343 & 10.47 & 0.048 (macro) & - & 3.28 \\
\hline
Supply Voltage (V) & 1.8 & 0.9 & 1.1 & 1.1 & 0.55--0.9 & 0.9 \& 0.29 \\
\hline
Input/Weight/Output bit & 1/4/1 & 1/4/1 & 1-8/1-4/1 & 4/4/4 & 1/4/1 & 1/1.5/1 \\
\hline
CIM Macro Size & 2.18 Kb & 10.625 Kb & 19 Mb & 10 Kb & 110 KB & 1.27 Mb \\
\hline
Bit-cell & 6T SRAM & RRAM & DRAM & 8T SRAM & 6T SRAM & 8T SRAM \\
\hline
Clock rate (MHz) & - & 300 & 250 & 200 & 51--280 & 71 \\
\hline
Energy Efficiency (TOPS/W) & - & 10.146 & 0.03 (I=4b, W=4b) & 54.26 & - & \textbf{1181.42} \\
\hline
Energy Efficiency (pJ/SOP) & 0.53 & - & - & - & 3.78 @0.55 V & \textbf{0.647} \\
\hline
Energy/inference (nJ) & - & - & - & 720 & 3800 & \textbf{410/277.7} \\
\hline
Accuracy (\%) & 90.5 & 85.6 & 94.13 & 92.1 & 97 & 93.64/85.59 \\
\hline
Dataset/Class & MIT-BIH Arrhythmia & Ninapro & CIFAR-10 & CIFAR-10 & NMNIST & GSCD/CIFAR-10 \\
\hline
\end{tabular}
Note: Normalization formula= (TOPs/W × InputBitwidth × WeightBitwidth × $(Process/28\,nm)^2$). TOPS refers to the Peak TOPS of the hardware. 
\label{tab:comparison}
\end{table*}

\section{Experimental Results}
\label{sec:results}
\subsection{Chip Implementation and Measurement Results}
\label{sec:chip_results}
The prototype of the AI accelerator is implemented using TSMC's 28\,nm CMOS process, featuring 1.27\,Mb of on-chip SRAM cells along with 64 distributed sensors and regulators. The large-scale SRAM-CIM macro facilitates weight updates with reduced latency. The chip size is 3.28 mm². 
Fig.~\ref{fig:chip_micrograph} displays the chip micrograph. 
The measured shmoo plot and the output voltage of the regulators are shown in Fig.~\ref{fig:shmoo_plot}, which ranges from 0.28\,V to 0.3\,V during the CIM mode operation.  The test chip was fabricated via an educational multi-project wafer (MPW) shuttle, where a limited number of dies are available. Experimental characterization of three sample chips yielded consistent performance across all units. Furthermore, the design's robustness against process, voltage, and temperature (PVT) variations was verified through extensive Monte Carlo simulations.

For the performance of the CIM mode operations, the readout circuit and neural network are reconfigurable for CNN or SNN operations. As an SNN accelerator, it supports programmable timesteps ranging from 1 to 3, as a trade-off between throughput and accuracy. The AI accelerator demonstrates robust operation down to approximately 0.29\,V and lowers overall power consumption by 40\%. The SNN accelerator achieves a throughput of 20.972/9.64/3.21 TOPS for peak/one time-step/three time-step operations respectively, and access time is about 240.37\,$\mu$s. 

Table~\ref{tab:comparison} presents the performance benchmarks on different datasets. For the keyword spotting using the GSCD/12 dataset, the accelerator achieves 93.64\% (three time-step) and 91.17\% (one time-step) accuracy, with an energy efficiency of 1181.42/1772.13 (normalized) TOPS/W and an area efficiency of 7.24/10.86 (normalized) TOPS/mm². The normalized TOPS is relative to single bit operation, equivalent to $TOPS\times(Input Bitwidth)\times(Weight Bitwidth)$. To further evaluate the proposed accelerator's adaptability, this work also tested it on the CIFAR-10 dataset. As shown in the table, an accuracy of 85.59\% was achieved when mapping the whole model to the macro without weight update due to the very small model size. This accuracy could be improved for a larger size model based on our evaluation. This performance is achieved with the same energy-efficient subthreshold CIM operation, demonstrating the robustness of the underlying hardware. The reported accuracy values are derived from software inference with hardware-calibrated parameters. Except for the input encoding layer, the entire model is executed on hardware, with all timesteps and neuron accumulation operations performed on silicon.

The overhead of the current threshold generation is shown below. For current sensing and regulation, each sense amplifier consumes 25.2 $\mu$W, while $I_{TH}$  generator consumes $0.9 \mu$W. The power overhead including sensing and threshold settings is about 0.9$\%$ of the total chip power.
Regarding area-overhead, each $I_{TH}$ block is implemented using five SRAM cells. For 128 instances, the total area overhead is about 364 $\mu$$m^2$.  It corresponds to only 0.0011$\%$ of the total chip area (3.28 $mm^2$), showing that the area and power overhead of the proposed current threshold generation scheme are actually negligible.

\subsection{Comparison to Other Works}
Table~\ref{tab:comparison} also presents a detailed comparison with other state-of-the-art SNN CIM accelerators. While prior works have made significant contributions, this work distinguishes itself through a superior combination of very high energy efficiency and robust operation implemented in a 28\,nm CMOS technology. For instance, compared to the RRAM-based design by~\cite{Tian2022} which operates at 0.9\,V, our work achieves a substantially higher normalized energy efficiency (1181.42~TOPS/W for one time-step vs. 10.146~TOPS/W) by successfully leveraging subthreshold (0.29\,V for SRAM array) operation. The eDRAM approach by~\cite{Kim2023}, while achieving high accuracy on CIFAR-10, reports a significantly lower energy efficiency of 0.03~TOPS/W (for 4-bit input/weight) and operates at a higher voltage of 1.1\,V. The 8T SRAM design in~\cite{Kim2023b}, also in 28\,nm, demonstrates ADC-less SNN operation but at 1.1\,V, achieving 54.26~TOPS/W; our work operates its 8T SRAM array at a much lower 0.29\,V, which is a key factor contributing to its order-of-magnitude improvement in energy efficiency. Even the recent advanced-node (22\,nm) SNN processor by~\cite{Liu2024}, which showcases impressive ultra-low synapse energy (0.26\,nJ/synapse, translating to 3.78~pJ/SOP at 0.55\,V) and high accuracy on NMNIST, targets a different optimization point (time-step-first dataflow for sparsity) and its system-level TOPS/W is not directly compared in the same manner, though our pJ/SOP (0.647~pJ/SOP) is also highly competitive. The work by~\cite{Liu2022}, using 6T SRAM in 180\,nm, achieves 0.53~pJ/SOP but at a higher supply of 1.8\,V.

The superior energy efficiency of our work primarily stems from the successful implementation of large-scale (1.27\,Mb macro) subthreshold current-mode CIM. This challenging operating regime was made possible by the novel integration of in-situ PVT compensation through monitor sensors and distributed regulators, allowing aggressive voltage scaling down to 0.29\,V for the SRAM array during CIM mode, which drastically reduces both dynamic and static power consumption. Furthermore, the architectural co-design, including the stride-tick batching schedule and pipelined pooling, contributes to efficient multi-time-step SNN processing without excessive intermediate storage overhead. Crucially, the variation-aware training methodology ensures that high inference accuracy (93.64\% on GSCD, 85.59\% on CIFAR-10) is maintained despite these aggressive circuit techniques and the inherent variability of subthreshold operation. While some specialized designs might achieve higher peak clock rates or marginally higher accuracy on specific datasets with different network models, our design demonstrates a compelling overall balance of high accuracy, exceptionally high energy efficiency, and robust operation on a mature CMOS process, making it highly suitable for energy-constrained edge AI applications.

\section{Conclusion}
\label{sec:conclusion}

This work introduced a PVT-resilient, subthreshold SRAM-based CIM macro for energy-efficient SNN inference. The proposed architecture integrates in-situ current sensors and distributed voltage regulators, enabling robust current-mode operation under wide PVT variations. A hardware-software co-design approach combines normalization-free convolution blocks, stride-tick batching, and programmable neuron thresholds, significantly improving efficiency and scalability. Furthermore, a variation-aware training methodology mitigates SRAM cell mismatches and sense amplifier offsets without incurring hardware overhead.
Fabricated in 28\,nm CMOS, the prototype achieves 93.64\% accuracy on a keyword spotting task and delivers up to 1181.42 TOPS/W, establishing a new benchmark for subthreshold SRAM-based CIM systems. These results highlight the potential of co-optimized architectures and training strategies to advance ultra-low-power, large-scale neuromorphic computing.

\bibliographystyle{IEEEtran}
\bibliography{references} %

\begin{IEEEbiography}[{\includegraphics[width=1in,height=1.25in,clip,keepaspectratio]{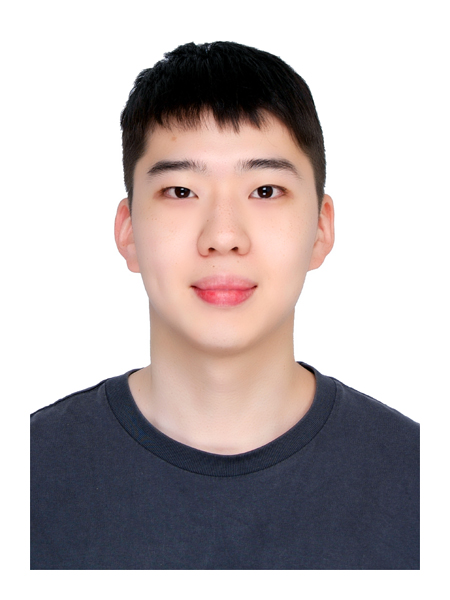}}]{Shih-Hang Kao}
received the B. S. and M. S. degrees in electronics engineering from National Yang Ming Chiao Tung University, Hsinchu, Taiwan, in 2021 and 2024. He is currently working at Realtek, Hsinchu, Taiwan. His research interest includes mixed-signal integrated circuit design and CIM.

\end{IEEEbiography}
\vspace{-11mm}
\begin{IEEEbiography}[{\includegraphics[width=1in,height=1.25in,clip,keepaspectratio]{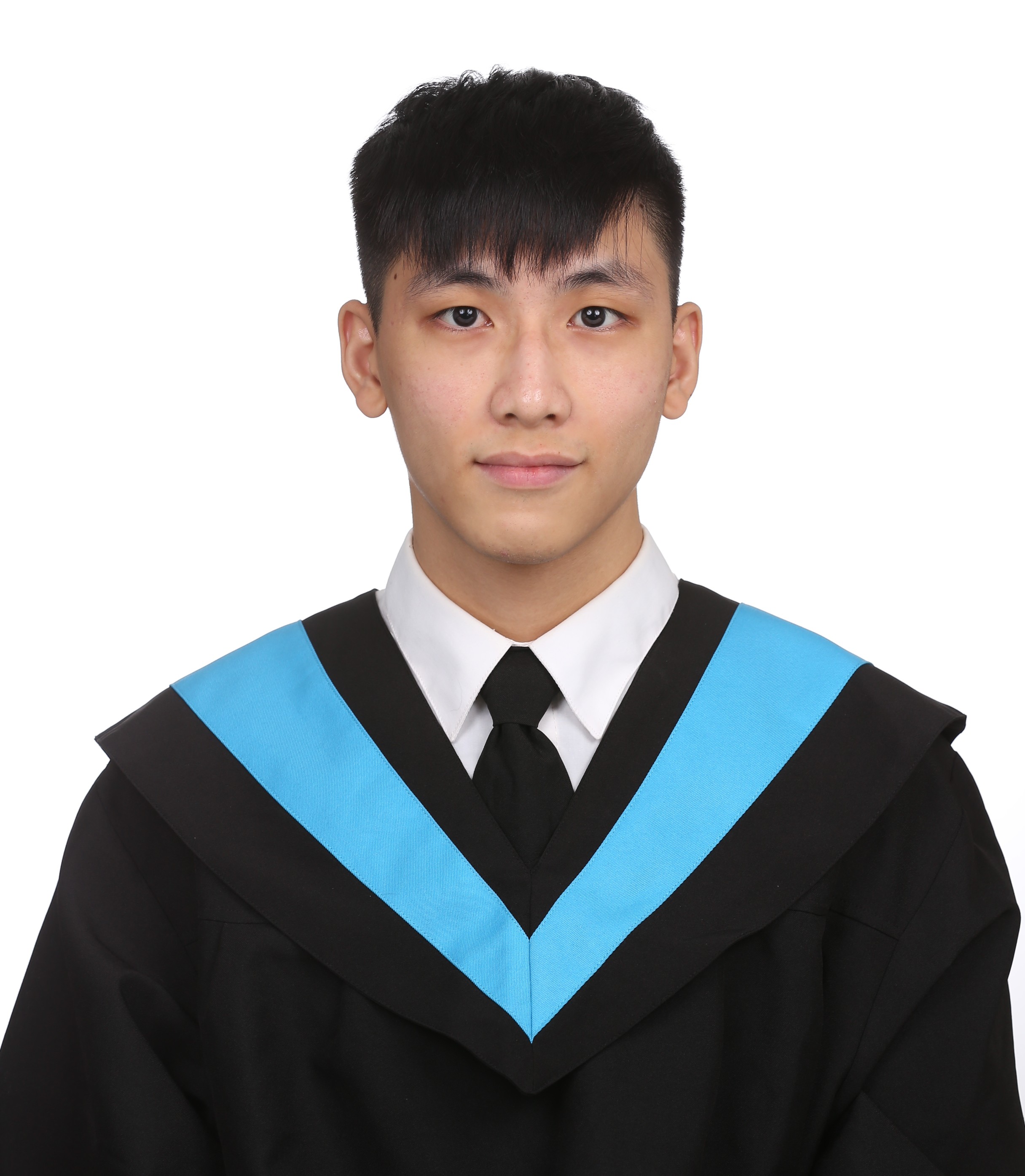}}]{Yang-Chan Hung} 
received the B. S. and M. S. degrees in electronics engineering from National Yang Ming Chiao Tung University, Hsinchu, Taiwan, in 2022 and 2024. He is currently working at NovaTek, Hsinchu, Taiwan. His research interest includes hybrid CIM and digital design and its mapping optimization for deep learning acceleration.

\end{IEEEbiography}
\vspace{-14mm}
\begin{IEEEbiography}[{\includegraphics[width=1in,height=1.25in,clip,keepaspectratio]{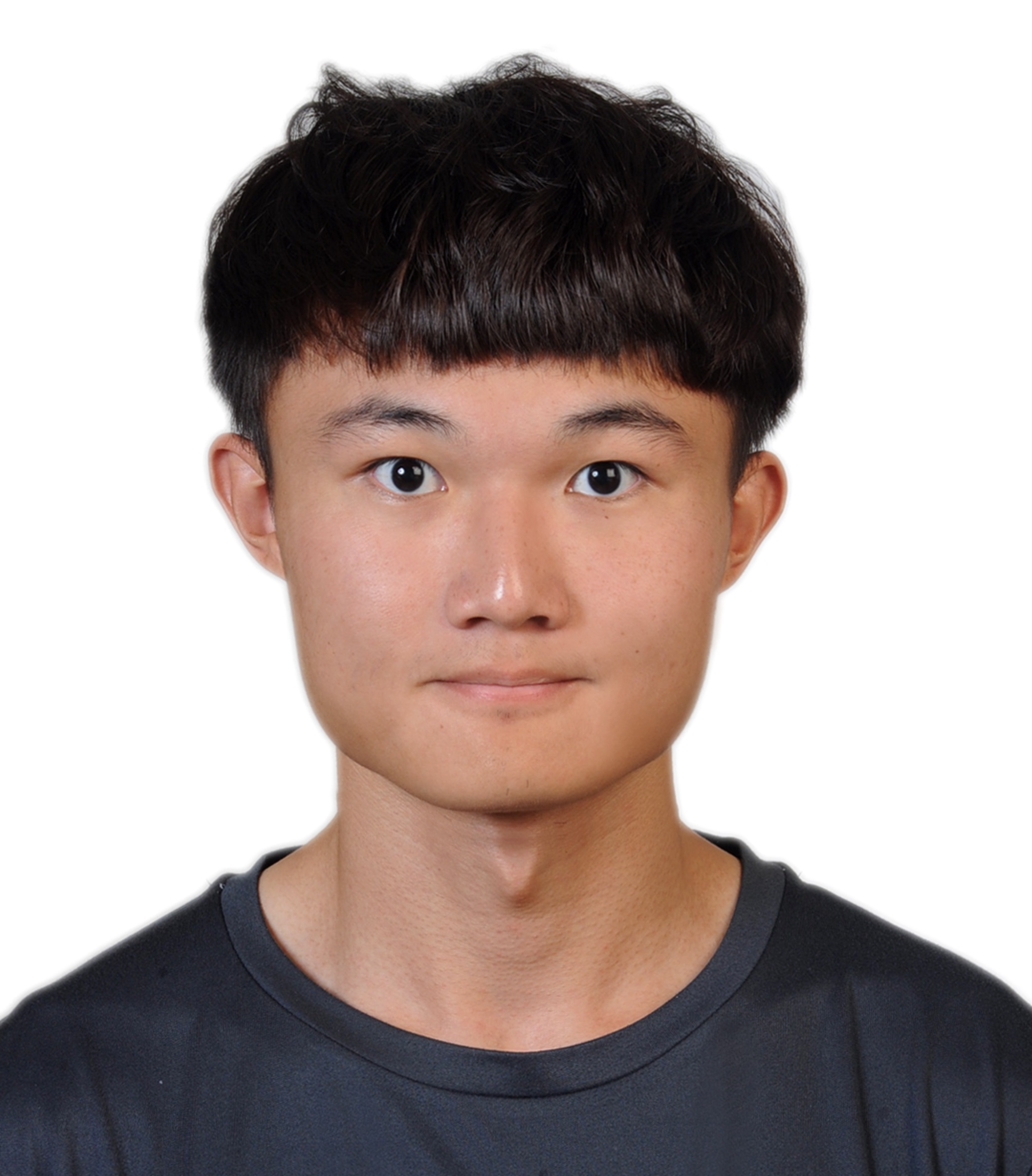}}]{I-Wen Wang}
received the M. S. degree in electronics engineering from National Yang Ming Chiao Tung University, Hsinchu, Taiwan, in 2023. He is currently working at Phison, Miaoli , Taiwan, engaged in digital circuit design for memory controller. His research interest includes SRAM in memory computing.

\end{IEEEbiography}
\vspace{-14mm}
\begin{IEEEbiography}[{\includegraphics[width=1in,height=1.25in,clip,keepaspectratio]{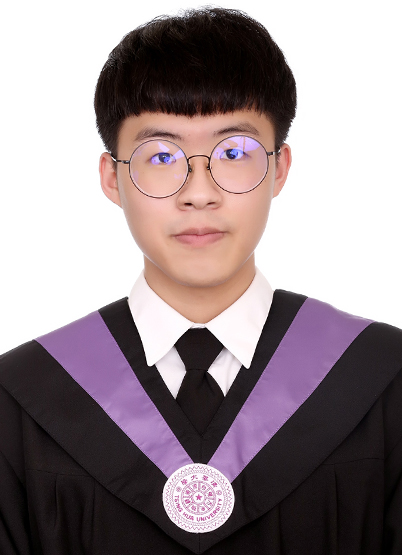}}]{Bing-Han Liu}
received the M. S. degree in institute of pioneer semiconductor innovation from National Yang Ming Chiao Tung University, Hsinchu, Taiwan, in 2024. He is currently working at Realtek, Hsinchu. His research interest includes SRAM in memory computing.
\end{IEEEbiography}
\vspace{-14mm}
\begin{IEEEbiography}[{\includegraphics[width=1in,height=1.25in,clip,keepaspectratio]{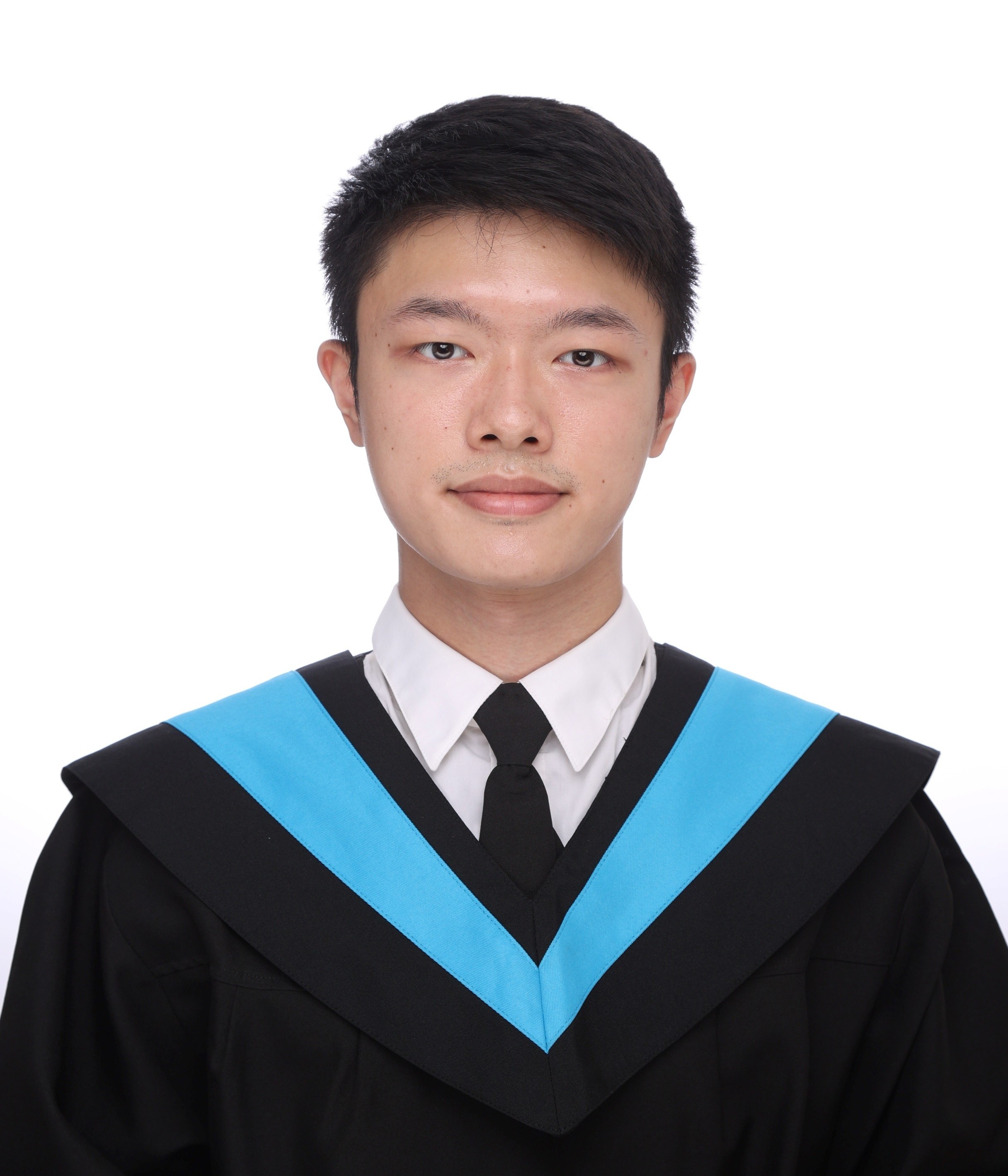}}]{Yu-Chia Chen}
received the B. S. degree in electronics and electrical engineering National Yang Ming Chiao Tung University, Hsinchu, Taiwan, in 2023. He is currently pursuing the M. S. degree in institute of pioneer semiconductor innovation from National Yang Ming Chiao Tung University, Hsinchu, Taiwan. His research interest includes mixed-signal integrated circuit design and CIM.
\end{IEEEbiography}

\vspace{-14mm}
\begin{IEEEbiography}[{\includegraphics[width=1in,height=1.25in,clip,keepaspectratio]{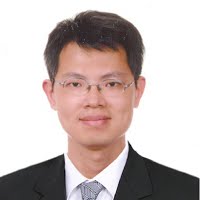}}]{Tian-Sheuan Chang}
	(S’93–M’06–SM’07)
	received the B. S., M. S., and Ph. D. degrees in electronic engineering from National Chiao-Tung University (NCTU), Hsinchu, Taiwan, in 1993, 1995, and 1999, respectively. 
	
	From 2000 to 2004, he was a Deputy Manager with Global Unichip Corporation, Hsinchu, Taiwan. In 2004, he joined the Department of Electronics Engineering, NCTU (as National Yang Ming Chiao Tung University (NYCU) in 2021), where he is currently a Professor. In 2009, he was a visiting scholar in IMEC, Belgium. His current research interests include system-on-a-chip design, VLSI signal processing, and computer architecture.
	
	Dr. Chang has received the Excellent Young Electrical Engineer from Chinese Institute of Electrical Engineering in 2007, and the Outstanding Young Scholar from Taiwan IC Design Society in 2010. He has been actively involved in many international conferences as an organizing committee or technical program committee member.
\end{IEEEbiography}
\vspace{-14mm}
\begin{IEEEbiography}[{\includegraphics[width=1in,height=1.25in,clip,keepaspectratio]{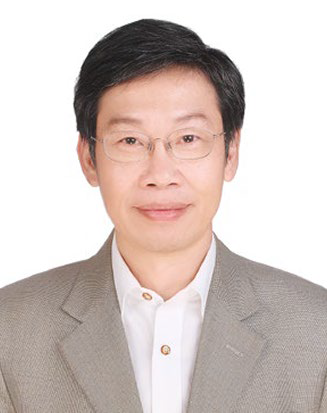}}]{Shyh-Jye Jou}
(Senior Member, IEEE) received the M. S. and Ph. D. degrees in electronics from National Chiao Tung University, Hsinchu, Taiwan, in 1984 and 1988, respectively.

From 1990 to 2004, he was with the Electrical Engineering Department, National Central University, Taoyuan, Taiwan, and became a Professor in 1997. Since 2004, he has been a Professor with the Electronics Engineering Department, National Chiao Tung University, and served as the Chairperson from 2006 to 2009. From January 2016 to December 2017, he was appointed as the Director General of the Science Education and International Cooperation Department, Ministry of Science and Technology, Taiwan.

He was a Visiting Research Professor with the Coordinated Science Laboratory, University of Illinois at Urbana-Champaign, USA, during 1993–1994 and the 2010 academic year. In the summer of 2001, he was a Visiting Research Consultant with the Communication Circuits and Systems Research Laboratory, Agere Systems, Allentown, PA, USA.

He has published over 100 IEEE journal and conference papers. His research interests include high-speed, low-power mixed-signal integrated circuits, and communication and bio-electronics integrated circuits and systems.

Dr. Jou received the Outstanding Engineering Professor Award from the Chinese Institute of Engineers in 2011 and from the Chinese Institute of Electrical Engineering in 2013. He served as the Chapter Chair for the IEEE Circuits and Systems Society Taipei Chapter in 2006 and as a Guest Editor for the \textsc{IEEE Journal of Solid-State Circuits} in November 2008. He has held various leadership roles in IEEE conferences, including serving as Conference Chair and TPC Chair for IEEE VLSI-DAT and the International Workshop on Memory Technology, Design, and Testing. He was also a Technical Program Chair or Co-Chair for IEEE VLSI-DAT, IEEE Asian Solid-State Circuits Conference, and IEEE Biomedical Circuits and Systems. He served as a Distinguished Lecturer for the IEEE Circuits and Systems Society from 2009 to 2010.
\end{IEEEbiography}
\vspace{-14mm}
\begin{IEEEbiography}[{\includegraphics[width=1in,height=1.25in,clip,keepaspectratio]{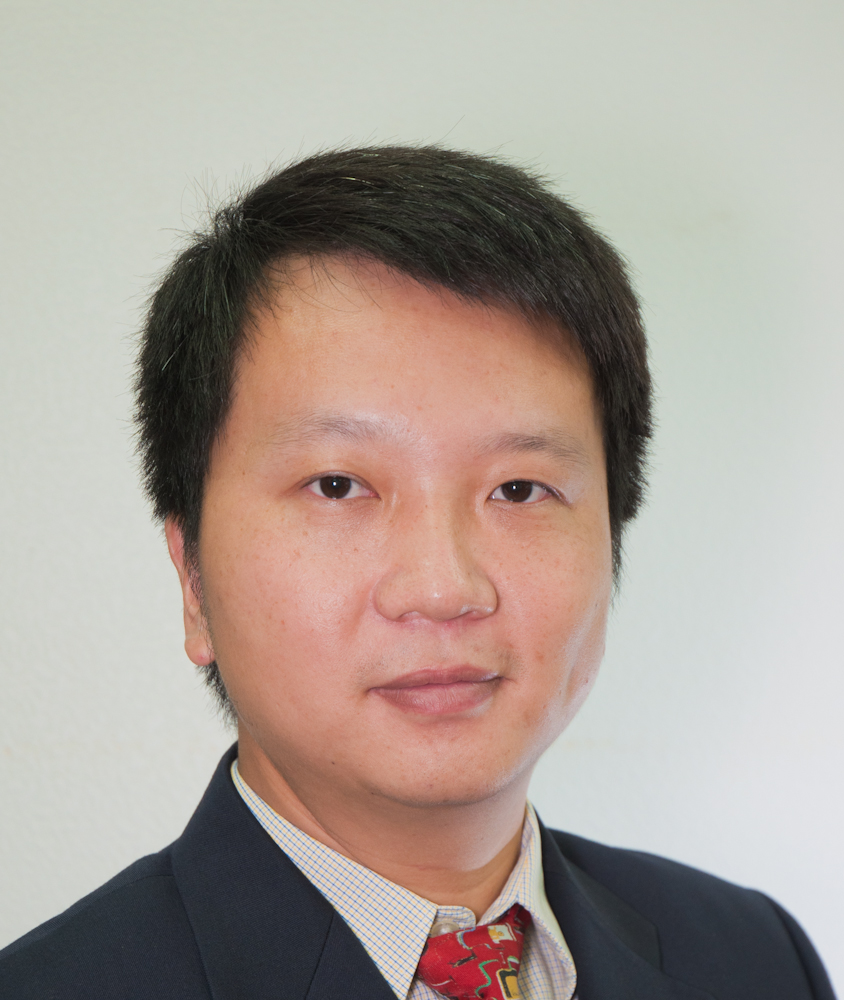}}]{Chien-Nan Liu}
	(M’98–SM’10)
	received the B. S. and Ph. D. degrees in electronics engineering from National Chiao Tung University, Hsinchu, Taiwan, in 1996 and 2001, respectively.
 
	He is currently a Full Professor of Institute of Electronics and the Deputy Director of Microelectronics and Information Systems Research Center, National Yang Ming Chiao Tung University. He has published more than 150 technical papers. His research interests include the behavioral modeling of analog designs, analog design automation techniques, mixed-signal verification techniques, and high-level power and noise modeling.
 
	Prof. Liu received the Outstanding Academy–Industry Cooperation Achievement Award granted by the Ministry of Education, Taiwan, in 2002, and the Outstanding Teaching Award granted by the National Central University, Taiwan, in 2006. He also received the Distinguished Young Scholar Award granted by the Taiwan IC Design Society in 2011. He has been elevated as a Senior Member of ACM in 2010. He is also a member of the Phi Tau Phi Scholastic Honor Society.
\end{IEEEbiography}
\vspace{-14mm}
\begin{IEEEbiography}[{\includegraphics[width=1in,height=1.25in,clip,keepaspectratio]{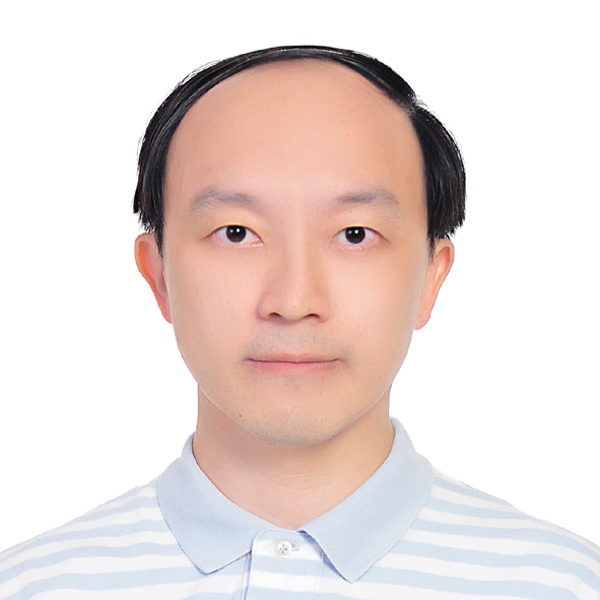}}]{Hung-Ming Chen}
	(M’03) received the B. S. degree in computer science and 
information engineering from National Chiao Tung University, 
Hsinchu, Taiwan, and the M. S. and the Ph. D. degrees in computer 
sciences from University of Texas at Austin.

Dr. Chen is currently a Professor at the Institute of 
Electronics at National Yang Ming Chiao Tung University, 
Hsinchu, Taiwan. He has served as the technical program 
committee members including ACM/IEEE DAC, ASP-DAC, IEEE/ACM ICCAD
and ACM ISPD. He also served as placement track chair in ICCAD 2023-25. He has supervised teams to win the first place at 2014 ISPD Placement Contest and 2023 CAD Contest at ICCAD. His research interests include design automation in digital and analog circuits, advanced packaging tool development and cell library layout generation in advanced technology.
\end{IEEEbiography} 
\vspace{-14mm}
\begin{IEEEbiography}[{\includegraphics[width=1in,height=1.25in,clip,keepaspectratio]{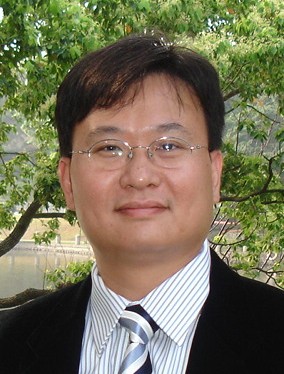}}]{Wei-Zen Chen}(M’00–SM’11) received the Ph. D. degree in electronics engineering from National Chiao Tung University (NCTU), Hsinchu, Taiwan in 1999. He is a professor at the Institute of Electronics and Department of Electronics and Electrical Engineering, National Yang Ming Chiao Tung University (NYCU). He served as the Institute director from 2015-2018. He was the Deputy Executive Director of the National SoC Program and the Principal Investigator of the National Project on Intelligent Electronics in Taiwan. He was an associate editor of the IEEE Solid-State Circuits Letters, a Guest Editor of the IEEE Journal of Solid-State Circuits, and served as a TPC member at IEEE International Solid-State Circuits Conference (ISSCC), IEEE Asian Solid-State Circuit Conference (A-SSCC) and IEEE Custom Integrated Circuits Conference (CICC). He was the IEEE Solid-State Circuits Society Taipei Chapter chair during 2008–2012, and is currently the chair of ISSCC APAC regional committee. He is a member of Phi Tau Phi Honorary Scholar society. His current research interests include mixed-signal integrated circuit for wireless and wireline communication systems, with a special emphasis on high-speed interconnects, optical communication, CIM, and radar sensing systems.

\end{IEEEbiography}

\end{document}